\documentclass[journal]{data/IEEEtran}
\ifCLASSINFOpdf
\else
\fi

\usepackage{color}
\usepackage{soul}
\usepackage[ruled]{algorithm}
\usepackage[noend]{algpseudocode}
\usepackage{algpseudocode}
\usepackage{listings}
\usepackage{fancyvrb}
\usepackage{xcolor}
\usepackage{graphicx}
\usepackage{float}
\usepackage{textcomp}
\usepackage{multirow}
\usepackage{array}
\usepackage{amsmath}
\usepackage{oubraces}
\usepackage{mathtools}
\usepackage{cite}
\usepackage{xspace}
\usepackage{subcaption}

\usepackage{hyperref}
\newcolumntype{C}[1]{>{\centering\let\newline\\\arraybackslash\hspace{0pt}}m{#1}}

\newcommand{\myequation}{\begin{equation}}
\newcommand{\myendequation}{\end{equation}}
\newcommand{\float}{IEEE-754\xspace}
\newcommand{\posit}{\emph{posit}\xspace}
\newcommand{\Posit}{\emph{Posit}\xspace}
\newcommand{\posits}{\emph{posits}\xspace}
\newcommand{\instbit}[1]{\mbox{\scriptsize #1}}    
\newcommand{\cclass}{C-class\xspace}

\newcommand{\instbitrange}[2]{~\instbit{#1} \hfill \instbit{#2}~}                                   
  
\newcolumntype{K}{>{\centering\arraybackslash}p{2.00cm}}
\newcolumntype{E}{>{\centering\arraybackslash}p{1.44cm}}

\begin{document}
%
\title{PERI: A Posit Enabled RISC-V Core}
%
%
%
\author{Sugandha Tiwari$^{\top}$, Neel Gala$^{\bot}$, Chester Rebeiro$^{\top}$ and V. Kamakoti$^{\top}$\\
$^{\top}$ Indian Institute of Technology Madras\\
$^{\bot}$ InCore Semiconductors Pvt. Ltd.\\
$^{\top}$ \{cs17s001, chester, kama\}@cse.iitm.ac.in, $^{\bot}$ neelgala@incoresemi.com
}

\maketitle

\begin{abstract}
Owing to the failure of Dennard's scaling the last decade has seen a steep growth of prominent new paradigms leveraging opportunities in computer architecture. Two technologies of interest are {\bf Posit} and {\bf RISC-V}.
Posit was introduced in mid-2017 as a viable alternative to IEEE 754-2008. Posit promises
more accuracy, higher dynamic range and fewer unused states along with simpler hardware
designs as compared to IEEE 754-2008. RISC-V, on the other hand, provides a commercial-grade  open-source
ISA. It is not only elegant and simple but also highly extensible and customizable, thereby facilitating novel micro-architectural research and exploration. In this paper, 
we bring these two technologies together and propose the first \textbf{Posit Enabled RISC-V} core.
The paper provides insights on how the current 'F' extension and the custom op-code space of RISC-V
can be leveraged/modified to support Posit arithmetic. We also present implementation details
of a parameterized and feature-complete Posit FPU which is integrated with the 
RISC-V compliant SHAKTI C-class core either as an execution unit or as an accelerator. 
To fully leverage the potential of Posit, we further enhance our Posit FPU, with minimal
overheads, to support two different exponent sizes (with posit-size being 32-bits). This allows
applications to switch from high-accuracy computation mode to a mode with higher dynamic-range at run-time.
In the absence of viable software tool-chain to enable porting of applications in the Posit domain, we present
a workaround on how certain applications can be modified minimally to exploit the existing RISC-V tool-chain.
We also provide examples of applications which can perform better with Posit as compared to IEEE 754-2008.
The proposed Posit FPU consumes 3507 slice LUTs and 1294 slice registers on an Artix-7-100T Xilinx FPGA while capable of operating at 100 MHz.

\end{abstract}

%
\IEEEpeerreviewmaketitle


\section{Introduction}
\label{sec:introduction}

For years, computer architects have cruised Moore's law~\cite{moore} and Dennard's 
scaling~\cite{dennard} to leverage increased resources to improve the quality of service of
general-purpose processors. However, recent explosion in domain-specific applications and the 
need for customized hardware has challenged architects to re-visit basic computer architecture 
principles for hidden opportunities. One such opportunity lies in computing with floating-point
numbers.

The \float standard~\cite{ieee754} has been considered the de facto floating-point standard for several decades. Hardware implementations catering to this standard are ubiquitous today in all 
major computing platforms. While it continues to drive many modern day applications, there 
exist quite a few acknowledged barriers in the \float standard, which has forced researchers
to find alternatives. For example, it is a well-known fact that the corresponding hardware
units require significant amount of area and energy of the chip~\cite{luca, fang}. 
Moreover, the complexity and inconsistencies within
the \float standard have led to several implementation errors of FPU (Floating Point Units)~\cite{whitehead, software_bugs}.
This has made compliance with the standard a major effort.
\float also suffers from lack of guaranteed re-producibility for some operations and across platforms as it recommends that compilers should ensure the same~\cite{ieee754}. The support for overflow and underflow further leads to loss of accuracy. 
The standard, even today,
allows multiple and redundant representations of NaNs (Not-a-Number) making software portability
across implementations a major challenge. While subnormals are rare-to-occur numbers, the
corresponding hardware to handle them adds significant overheads to the entire design. Though today, the standard elaborates on
16-bit, 32-bit, 64-bit, 128-bit, and 256-bit representations, each bit-width can support only 
fixed precision and dynamic range. Despite these challenges, the lack of a suitable replacement 
has forced architects to continue with the \float standard for floating-point numbers.


\par Quite recently, \posits~\cite{posit} have been proposed as an alternate and efficient way of
representing floating-point numbers. Unlike \float, \posits enable reproducible results across
platforms. By not supporting overflow and underflow, \posits preserve the remaining information
by rounding down or up. \Posit based arithmetic is significantly more simplified due to the absence of
subnormals and NaNs, thereby also simplifying handling of exceptions to a large extent. 
Unlike \float, \posits are not restricted by a constant exponent and fraction size. 
A \posit representation is determined by the exponent size (\emph {es}) parameter, allowing one to select
different \emph{es} values to get different precision and dynamic range for the same posit-size ({\emph ps}).

Several works in literature~\cite{date, iscas, conga, ipdps, sigproc, iccd, pacogen} have proposed techniques
and methodologies to implement/generate hardware for \posit based arithmetic operations such as: adders, multipliers and dividers. 
Though these operations form an integral part of many \posit based FPUs (Floating Point Units), they are not sufficient to support
the compute requirements of a commercial-grade ISA (Instruction Set Architecture). Such a support would require functional units 
which can perform conversion of a number from the \posit domain to integer domain and vice-versa, compare \posit numbers, 
square-root operations, classification operations and much more. 
To that extent, in this paper we propose designs of relevant functional units required to build a complete \posit based FPU which can
support the floating-point compute extension of the RISC-V~\cite{riscv} ISA. Furthermore, the paper also describes methodologies 
of integrating the proposed \posit FPU with an open-source general-purpose RISC-V core either as a tightly-coupled execution unit within the core pipeline or as co-processor.
 
One of the major attributes of \posit is its ability to facilitate different dynamic range (and accuracy) for the same {\emph ps} value by simply 
manipulating the value of {\emph es}. Nonetheless,
all of the previous works on \posit have focused on designing/generating hardware compute blocks for a single
pair-value of {\emph es} and {\emph ps}. 
In this paper, we extend the functionality of our \posit
ALU to support two {\emph es} values ({\emph es} = 2 and 3), with minimal overheads. We propose to extend the 
control and status register to hold the current value of \emph{es} for all operations. Thereby, enabling the user to choose either 
higher accuracy or higher dynamic range for the same \emph{ps} size.

In the summary, the contributions of this paper are as follows:
\begin{enumerate}
    \item The paper provides insights on how the RISC-V ISA can be leveraged, modified and customized to support \posit based 
    compute. 
    \item Design and implementation details of all functional units required to build a RISC-V enabled 
    \posit FPU have been proposed. These implementations are parameterized for {\em (ps, es)} values and have been designed using Bluespec System Verilog.
    \item The proposed \posit FPU has been integrated with an open-source RISC-V core as both: a tightly-coupled execution unit and a co-processor. To the best of the authors' knowledge, this is the first work to provide a complete processor with \posit support.
    \item The \posit FPU has been further enhanced, with minimal overheads, to support two {\emph es} values in same HW, thereby enabling 
    dynamic switching between higher accuracy or higher dynamic range at run-time.
    \item The paper also presents analysis of various software applications running on the core, which provide similar or better performance
    in terms of quality as compared to \float.
    
\end{enumerate}

The rest of the paper is organized as follows. Section-\ref{sec:background} provides a brief
 introduction and background to the \posit format and the RISC-V ISA. Section-\ref{sec:isa-extension} discusses
 the details of extending the RISC-V ISA to accommodate \posit arithmetic operations. The entire \posit floating-point
 unit is discussed in detail in Section-\ref{sec:posit_arith}. Integration of the proposed \posit unit with a RISC-V
 core is highlighted in Section-\ref{sec:core-integration}. The software modifications required to port \posit based
 applications on a RISC-V are covered in Section-\ref{sec:sw_support}. Section-\ref{sec:applications} presents
 the results and insights obtained from running a few interesting applications on the \posit enabled RISC-V core. The hardware results are reviewed in Section-\ref{sec:hw-results}
 followed by brief literature survey in Section-\ref{sec:related_work}. Section-\ref{sec:conclusion} finally concludes the paper.
 

\section{Background}
\label{sec:background}

\subsection{The Posit Format}
\label{sec:posit_format}

The \posit representation is defined using two parameters: the \emph{ posit size (ps)} and \emph{exponent size (es)}. A formal representation of a \posit number is shown in Equation-\ref{eq:posit_format}.

\begin{equation}
\label{eq:posit_format}
\underbrace{
    \overbrace{s
    }^\text{\textit{Sign}} \;
    \overbrace{r \; r \; r \; ... \; r \; \overline{r}
    }^\text{\textit{Regime}} \;
    \overbrace{e_1 \; e_2 \; e_3 \; ... \; e_{es}
    }^\text{\textit{Exponent,if any}} \;
    \overbrace{f_1 \; f_2 \; f_3 \; f_4 \; ...
    }^\text{\textit{Fraction,if any}} \;
}_\text{posit size}
\end{equation}

 A \posit number utilizes the 2's complement notation to represent a negative number. The first bit (from the left) of the \posit number is the sign bit (denoted by \emph{s}). The number of identical bits following the sign bit, are called regime bits, and they determine the value of the variable \emph{k}. If there are {\em rc} identical bits in a number, then value of {\em k} is determined by Equation-\ref{eq:k_value}. The {\emph es} number of bits after the regime bits (denoted by \emph{e}) help determine the \posit exponent. The \posit exponent value (denoted by \emph{exp}) is a combination of regime and {\em e} bits and is derived using Equation-\ref{eq:exp}. The remaining bits trailing the {\emph e} bits, forms the fraction of the number. The implicit (hidden) bit is always 1 in case of \posit fraction (denoted by {\emph f}). Thus, the value \emph{x} of a \posit number \emph P is determined by Equation-\ref{eq:posit_value}. 
 
 As can be seen from Equation-\ref{eq:posit_value} the only two exceptions in \posit representations are 0 and \emph{NaR} (Not-a-Real) numbers. This makes exception handling very simple in \posit as compared to \float.
 
\begin{equation}
\label{eq:k_value}
    k  =
\begin{dcases}
    rc-1 & \text{if regime starts with 1} \\ 
    -rc  & \text{otherwise}
\end{dcases} 
\end{equation}

\begin{equation}
\label{eq:exp}
    exp = (k \ll es) + e
\end{equation}

\begin{equation}
\label{eq:posit_value}
    x  =
\begin{dcases}
    0 & P = 000 ... 000 \\
    \text{\textit{NaR}} & P=100 ... 000 \\ 
    (-1)^s \times 2^{exp} \times 1.f  & \text{otherwise}
\end{dcases} 
\end{equation}

Another significant difference between \posit and \float representation is that \posit uses run-length encoding to represent the exponent value. This enables a \posit number with a small value of \emph{es} to represent higher precision numbers than \float. Similarly, larger \emph{es} values can express a much higher dynamic range than \float.

\subsection{The RISC-V ISA}
\label{ssec:riscv_isa}

RISC-V (pronounced "risk-five") is a fairly new ISA, which has received tremendous momentum and support in the recent
years by both, academia and industry. RISC-V is a completely open ISA suitable for real HW implementations and not just
simulation or binary translation. The ISA has been well designed with a \emph{small}, but usable, 
base Integer ISA (RV32I) and does not mandate any particular micro-architecture style. 
The attribute of optional standard extensions and the ability to add custom ISA extensions has been the most attractive features of RISC-V. 

RISC-V today is supported by a strong and vibrant software ecosystem which includes support for: gcc, binutils, llvm, gdb, open-ocd, linux kernel, 
sel4 and much more. There have also been significant efforts world-wide to build open-source processors around RISC-V. Some of the prominent works 
include: SHAKTI\cite{shakti}, Rocket-Chip\cite{rocket}, lowRisc\cite{lowrisc} and Ariane\cite{ariane}. Commercial entities such as Western-Digital
and Bluespec Inc. have also contributed their core implementations (SweRV\cite{swerv} and Piccolo/Flute\cite{piccolo} respectively) to the
open-source community. 

With regards to floating-point compute, the RISC-V ISA includes two standard extensions: 'F' for single-precision
floating-point and 'D' for double-precision floating-point compute. Each of these extensions comprises of instructions
compliant with the \float arithmetic standard. The 'F' extension requires a separate 32-bit floating-point register file
while the support for 'D' extension requires a 64-bit floating-point register file. Though, the RISC-V specification
defines these extensions as \emph{standard extensions}, it also allows users to modify these extensions as per their
will, while still being compliant with other standard extensions of the specification. In the next section, we
leverage this opportunity and describe how the current 'F' extension can be modified to support \posit arithmetic.

Within the 32-bit instruction format, the ISA has also locked down on two major-opcodes
as \emph{custom}, meaning that these opcodes can be used to define user-specific 
custom instructions with a guarantee that no future standard extensions shall use this 
opcode space. Subsequent sections of this paper exploit this opportunity to present
how this custom opcode space can be leveraged to build a larger \posit based co-processor compliant with any RISC-V device.
\section{The RISC-V POSIT Extensions}
\label{sec:isa-extension}

This section describes how the RISC-V ISA can be leveraged and modified to include \posit based arithmetic
operations. We propose two separate approaches to ISA extensions. The first approach is to leverage the 'F'
extension of the ISA itself with minimal modifications to support \posit based arithmetic. This approach requires
none/minimal changes in the software tool-chain and thus enabling quick bring up of the design.

The second approach relies on utilizing the \emph{custom-0/1/2/3} major-opcodes space of the 32-bit instruction format. 
This opcode space has been frozen by the ISA to be used for integrating custom instruction sets and will be future compatible. This approach allows us to expand the \posit FPU capabilities beyond the operations specified
in the 'F' extension. Furthermore, this method enables integrating a \posit based FPU as a co-processor to any
RISC-V core which forwards all \emph{custom} opcodes externally using a standard co-processor interface. Unlike,
the previous approach, this approach does not require any modifications to occur within the core-pipeline and also facilitates the co-existence of the \float and \posit compute units on the same chip. 
However, utilizing the \emph{custom-0/1/2/3} opcodes requires changes to the software tool-chain.

Each of the above approaches is discussed in detail in the following subsections. 

\subsection{Leveraging the 'F' extension}
\label{sec:f-extension}
Before proceeding further, the reader is recommended to be cognizant with the 'F' extension of the RISC-V ISA.
A gist of instructions comprising the 'F' extension are captured in Table-\ref{tab:rv32f}. 
\begin{table}[!t]
    \small
    \begin{tabular}{|p{4.8cm}|C{2.8cm}|}
      \hline
      \textbf{Instructions} & \textbf{Description}  \\ \hline \hline
       FMADD.S, FMSUB.S, FNMSUB.S, & \multirow{3}{2.8cm}{\centering Fused-Multiply-Add ops} \\
       FNMADD.S, FADD.S, FSUB.S, & \\
        FMUL.S & \\
      \hline
      FDIV.S & Division op\\  \hline
      FSQRT.S & Square Root op\\  \hline
      FSGNJ.S, FSGNJN.S, FSGNJX.S & Sign Injection ops\\ \hline
      FMIN.S, FMAX.S, FEQ.S, FLT.S, FLE.S & {Comparison ops} \\ \hline
     
      FCVT.W.S, FCVT.WU.S, &
      \multirow{2}{2.8cm}{\centering Conversion ops} \\
      FCVT.S.W, FCVT.S.WU & \\ \hline
      FMV.X.W, FMV.W.X & Transfer ops \\ \hline
      FCLASS.S  & Classification op \\ \hline
      
    \end{tabular}
    \caption{List of instructions comprising the F extension of RISC-V.}
    \label{tab:rv32f}
\end{table}

\begin{figure}[!t]
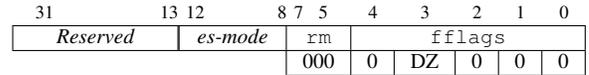
                                                                                  
{\footnotesize                                                                                      
\begin{center}                                                                                      
\begin{tabular}{K@{}E@{}cccccc}
\instbitrange{31}{13} &
\instbitrange{12}{8} &
\instbitrange{7}{5} &
\instbit{4} &
\instbit{3} &
\instbit{2} &
\instbit{1} &
\instbit{0} \\
\hline      
\multicolumn{1}{|c|}{{\em Reserved}} &
\multicolumn{1}{|c|}{{\em es-mode}} &
\multicolumn{1}{c|}{\tt rm} &
\multicolumn{5}{c|}{{\tt fflags}} \\
\hline
\multicolumn{1}{c}{} &
\multicolumn{1}{c|}{} &
\multicolumn{1}{c|}{000} &
\multicolumn{1}{c|}{0} &
\multicolumn{1}{c|}{DZ} &
\multicolumn{1}{c|}{0} &
\multicolumn{1}{c|}{0} &
\multicolumn{1}{c|}{0} \\
\cline{3-8} \\
\end{tabular}
\end{center}
}                                                                                                   
\vspace{-0.1in}                                                                                     
\caption{Posit control and status register (pcsr).}                                               
\label{fig:pcsr}    
\end{figure}

We propose to maintain the same register-file state for \posit as that of the 'F' extension, i.e. 32 \posit registers: 
{\bf p0-p31} each 32-bits wide. The \posit variant of the control and status register is shown in 
Figure-\ref{fig:pcsr}. Since \posit support only one rounding mode (round-to-nearest with tie-to-even), 
the \emph{rounding mode (rm)} field is not required and thus tied to zeros. Similarly, there
is no requirement of flags for invalid, inexact, overflow and underflow. The \posit exception of 
NaR (Not-a-Real) is silent and thus does not get captured in the flags. 
The exception of divide-by-zero is mapped to the {\bf DZ} field of \emph{fflags}. 
The \emph{pcsr} register also holds a 5-bit \emph{es-mode}
field, which indicates the current value of \emph{es} being used by the \posit FPU to deduce the \posit number. To maintain compatibility with the 'F' extension, the \emph{ps} value is set to 32-bits, and we expect all practical implementations
of a \posit FPU to support \emph{es} values which can be represented within the 5-bit field.
While the majority of implementations would support only a single value of \emph{es} (thus causing this field
to be read-only), later parts of this paper propose a \posit FPU design that can support up to two different \emph{es} values thereby using the \emph{es-mode} field to perform the switching. This field can be modified
using the standard CSR instructions.
Implementations which support multiple \emph{es-mode} values, should rely on the software to perform
a probe-and-find mechanism to identify all legal \emph{es} values supported by the platform.

Regarding \float, the RISC-V spec mandates that any floating-point operation resulting in a NaN 
(Not-a-Number) should output a canonical NaN (i.e. {\tt 0x7fc00000}). However, in \posit, NaR
has a single representation which maps to the most negatives 2's complement signed integer. The fact
that there is no notion of 'unorderedness' in \posits, allows a user to leverage integer-based comparison
techniques to compare \posit numbers.

All instructions proposed in the 'F' extension behave the same way for \posits as they do for
\float. The encoding of all instructions remains the same. The \emph{rm} field in all instructions, 
except the posit-to-integer conversion ops, is ignored since \posit only supports a single rounding
mode. For the posit-to-integer conversion ops, we realized that by supporting the round-to-zero mode, certain
applications, like JPEG compression, are able to provide much better results as compared to using the default
round-to-nearest mode. Thus, we propose to keep the \emph{rm} field in these instructions to mean the
same as they do in the default spec.

Since \posit does not have subnormals, different types of NaN values, infinities or different kind
of zeros, the classify operation will only capture if the operand is a zero, NaR, negative or positive,
leaving the other bits to be zeros always.

\begin{table*}[!tbp]                                                                                  
{\footnotesize                                                                                      
\begin{center}                                                                                      
\begin{tabular}{E@{}E@{}cccccccc}
\instbitrange{31}{25} &
\instbitrange{24}{20} &
\instbitrange{19}{15} &
\instbit{14} &
\instbit{13} &
\instbit{12} &
\instbitrange{11}{7} &
\instbitrange{6}{0} \\
\hline      
\multicolumn{1}{|c|}{fn} &
\multicolumn{1}{c|}{rs2} &
\multicolumn{1}{c|}{rs1} &
\multicolumn{1}{c|}{xd} &
\multicolumn{1}{c|}{xs1} &
\multicolumn{1}{c|}{xs2} &
\multicolumn{1}{c|}{rd} &
\multicolumn{1}{c|}{op} \\
\hline

\end{tabular}
\end{center}
}                                                                                                   
\vspace{-0.1in}                                                                                     
\caption{RoCC interface standard instruction format.}                                               
\label{fig:rocc-instruction}    

\begin{small}
\begin{center}
\begin{tabular}{p{0in}p{0.4in}p{0.05in}p{0.05in}p{0.05in}p{0.05in}p{0.4in}p{0.6in}p{0.4in}p{0.6in}p{0.7in}l}
& & & & & & & & & & \\
                      &
\multicolumn{1}{l}{\instbit{31}} &
\multicolumn{1}{r}{\instbit{27}} &
\instbit{26} &
\instbit{25} &
\multicolumn{1}{l}{\instbit{24}} &
\multicolumn{1}{r}{\instbit{20}} &
\instbitrange{19}{15} &
\instbitrange{14}{12} &
\instbitrange{11}{7} &
\instbitrange{6}{0} \\
\cline{2-11}

&
\multicolumn{4}{|c|}{funct7} &
\multicolumn{2}{c|}{rs2} &
\multicolumn{1}{c|}{rs1} &
\multicolumn{1}{c|}{funct3} &
\multicolumn{1}{c|}{rd} &
\multicolumn{1}{c|}{opcode} & R-type \\
\cline{2-11}

&
\multicolumn{2}{|c|}{rs3} &
\multicolumn{2}{c|}{funct2} &
\multicolumn{2}{c|}{rs2} &
\multicolumn{1}{c|}{rs1} &
\multicolumn{1}{c|}{funct3} &
\multicolumn{1}{c|}{rd} &
\multicolumn{1}{c|}{opcode} & R4-type \\
\cline{2-11}

&
\multicolumn{6}{|c|}{imm[11:0]} &
\multicolumn{1}{c|}{rs1} &
\multicolumn{1}{c|}{funct3} &
\multicolumn{1}{c|}{rd} &
\multicolumn{1}{c|}{opcode} & I-type \\
\cline{2-11}

&
\multicolumn{4}{|c|}{imm[11:5]} &
\multicolumn{2}{c|}{rs2} &
\multicolumn{1}{c|}{rs1} &
\multicolumn{1}{c|}{funct3} &
\multicolumn{1}{c|}{imm[4:0]} &
\multicolumn{1}{c|}{opcode} & S-type \\
\cline{2-11}

\end{tabular}
\end{center}
\end{small}
\caption{RISC-V 32-bit instruction types used in RV32F}
\label{tab:rv32f-types}

\begin{small}
\begin{center}
\begin{tabular}{p{0in}p{0.4in}p{0.05in}p{0.05in}p{0.05in}p{0.05in}p{0.4in}p{0.6in}p{0.1in}p{0.1in}p{0.1in}p{0.6in}p{0.7in}l}
& & & & & & & & & & & & \\
                      &
\multicolumn{1}{l}{\instbit{31}} &
\multicolumn{1}{r}{\instbit{27}} &
\instbit{26} &
\instbit{25} &
\multicolumn{1}{l}{\instbit{24}} &
\multicolumn{1}{r}{\instbit{20}} &
\instbitrange{19}{15} &
\instbit{14} &
\instbit{13} &
\instbit{12} &
\instbitrange{11}{7} &
\instbitrange{6}{0} \\
\cline{2-13}

&
\multicolumn{4}{|c|}{funct7} &
\multicolumn{2}{c|}{rs2} &
\multicolumn{1}{c|}{rs1} &
\multicolumn{1}{c|}{xd} &
\multicolumn{1}{c|}{xs1} &
\multicolumn{1}{c|}{xs2} &
\multicolumn{1}{c|}{rd} &
\multicolumn{1}{c|}{custom-0/1/2/3} & R-type \\
\cline{2-13}

&
\multicolumn{2}{|c|}{rs3} &
\multicolumn{2}{c|}{funct2} &
\multicolumn{2}{c|}{rs2} &
\multicolumn{1}{c|}{rs1} &
\multicolumn{1}{c|}{xd} &
\multicolumn{1}{c|}{xs1} &
\multicolumn{1}{c|}{xs2} &
\multicolumn{1}{c|}{rd} &
\multicolumn{1}{c|}{custom-0/1/2/3} & R4-type \\
\cline{2-13}

&
\multicolumn{6}{|c|}{imm[11:0]} &
\multicolumn{1}{c|}{rs1} &
\multicolumn{1}{c|}{xd} &
\multicolumn{1}{c|}{xs1} &
\multicolumn{1}{c|}{funct1} &
\multicolumn{1}{c|}{rd} &
\multicolumn{1}{c|}{custom-0/1/2/3} & I-type \\
\cline{2-13}

&
\multicolumn{4}{|c|}{imm[11:5]} &
\multicolumn{2}{c|}{rs2} &
\multicolumn{1}{c|}{rs1} &
\multicolumn{1}{c|}{funct1} &
\multicolumn{1}{c|}{xs1} &
\multicolumn{1}{c|}{xs2} &
\multicolumn{1}{c|}{imm[4:0]} &
\multicolumn{1}{c|}{custom-0/1/2/3} & S-type \\
\cline{2-13}

\end{tabular}
\end{center}
\end{small}
\caption{Equivalent mappings of Table-\ref{tab:rv32f-types} instruction formats to RoCC instruction format.}
\label{tab:rv32p-types}
\end{table*}

\subsection{Leveraging the 'Custom' opcode space}
\label{sec:custom-isa}

While the F extension includes a nice subset of standard floating-point instructions which can cater
to a wide variety of applications, more often than not, there is always a need to extend ISA to add
more complex functionalities. Given the fact that
the 'F' extension is reserved and cannot be extended or modified for custom use, one would have to resort to the custom opcode space of RISC-V to extend ISA support. Moreover, as the complexity continues to grow, it would seem beneficial to integrate such an FPU as an accelerator rather than a tightly
coupled execution unit within the core pipeline. This reduces the risk of modifying the pipeline of the core, and more importantly, enables a stand-alone \posit FPU design which can be integrated to any other RISC-V core
which adheres to a standard co-processor interface. 

In this paper, we choose to adopt the RoCC~\cite{rocc} (Rocket Custom Co-processor) interface for our \posit FPU co-processor. 
All RoCC compliant accelerators follow a standard instruction format, as shown in Table-\ref{fig:rocc-instruction}. 
The \emph{xs1}, \emph{xs2}, and \emph{xd} bits control how the base integer registers are read and written by the
accelerator instructions. If the \emph{xs1/xs2} bit is set, then the corresponding integer register specified by rs1/rs2 is passed on
to the accelerator. If the \emph{xs1/xs2} bit is clear, then the accelerator can either re-use the \emph{rs1/rs2} field to encode
other information or use it to access its own register-file (the posit register file in our case). Setting the \emph{xd} fields operate in a similar manner but performs writes instead of reads on the respective register. 
RoCC not only enables the transfer of register-file data but also equips the accelerator with a memory-interface. We now
discuss how the 'F' extension can be mapped to the standard instruction format of Table-\ref{fig:rocc-instruction}
using the custom opcode spaces: \emph{custom-0/1/2/3}.

Table-\ref{tab:rv32f-types} shows the four instruction formats used by the
default 'F' extension. The 'F' extension uses: 9 I-type, 1 S-type, 4 R4-type and 12 R-type instructions. 
As discussed earlier, \posits support only a single rounding mode and thus the \emph{rm} field of the particular
instructions can be ignored completely. This introduces more bits to encode information.
Table-\ref{tab:rv32p-types} shows equivalent mappings of I, S, R4 and R-type instruction formats to a RoCC
based instruction format which can be leveraged by the \posit based FPU. We have not provided
a separate \emph{xs3} field since for all floating operations
the rs3 value is always read from floating register-file. A quick observation reveals that
the custom space for \posit can accommodate up to 512 R-type and 16 R4-type unique instructions. 
Additionally, from the I-type format, one can either have up to 8 instructions which utilize immediate and rs1
or can have up to 32K single-operand instructions (i.e. only rs1 is used. E.g. FSQRT). Similarly, one can implement
8 unique store-like instructions employing the S-type format. Later sections of this paper, will provide a 
brief overview of how these instructions are fetched by the core processor and offloaded to the accelerator for computing. 

At this point, the authors would like to highlight the fact that none of the proposed approaches have any significant
impact on the arithmetic compute units of the \posit based FPU. Both of the above-discussed approaches only impact
the integration scheme of an FPU with a core and have a close-to-none impact on the decoder. All the arguments regarding
NaR, flags, es-mode, etc made in the previous sub-section hold in this solution as well.

\section{Posit FPU}
\label{sec:posit_arith}

\begin{figure*}[t]
    \centering
    \includegraphics[width=\linewidth]{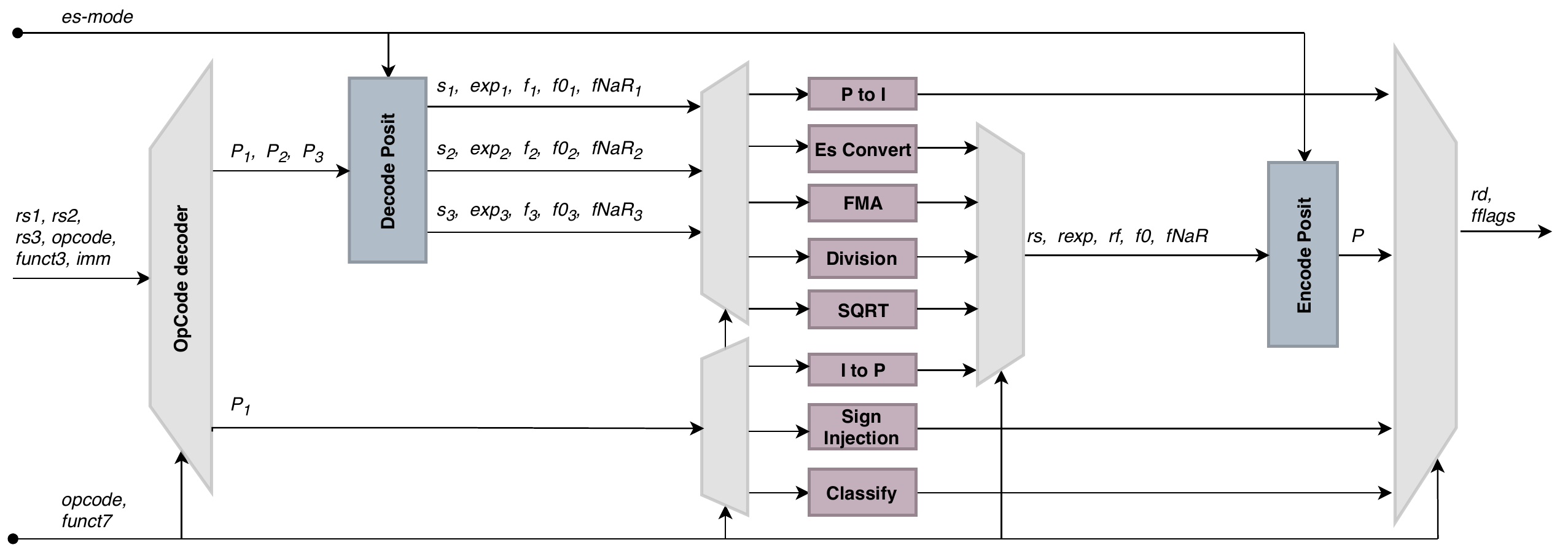}
    \caption{Block diagram showing the different components of \Posit unit.}
    \label{fig:posit_unit}
\end{figure*}

This section describes the various components of a \posit FPU for a RISC-V processor. We have implemented our design using Bluespec System Verilog (BSV). Our \posit FPU is parameterized to generate hardware for any combination of {\em ps} and {\em es} value. Figure-\ref{fig:posit_unit} shows the various components present in our implementation. Our \posit FPU has the following BSV interface definition. The input parameters of the interface refer to the fields of Table-\ref{tab:rv32f-types} and \ref{tab:rv32p-types}. \\

\begin{Verbatim}[commandchars=\\\{\},frame=single,fontsize=\footnotesize ,samepage=true]
interface Ifc_fpu;
  method Action _start(
    Bit#(32) rs1, Bit#(32) rs2, Bit#(32) rs3,
    Bit#(4)  opcode, Bit#(7) funct7,
    Bit#(3) funct3, Bit#(2) imm, Bit#(5) es-mode);
  method ActionValue#(Bit#(32)) get_rd;
  method ActionValue#(Bit#(5)) get_fflags;
endinterface
\end{Verbatim}

 \begin{algorithm}[!t]
    \caption{Algorithm for \Posit Decoding}
    \label{algo:decode_posit}
    \begin{algorithmic}[1]
        \renewcommand{\algorithmicrequire}{\textbf{Input:}}
        \renewcommand{\algorithmicensure}{\textbf{Output:}}
        \Require {\bf \em P:} floating-point number in posit format with $ps$ bits
        \Ensure {\bf \em s:} sign of {\bf \em P}, {\bf \em exp:} final exponent of {\bf \em P}, {\bf \em f:} fraction bits of {\bf \em P} including hidden bit, {\bf \em f0:} set if {\bf \em P} is 0, {\bf \em fNaR:} set if {\bf \em P} is NaR \\
        {\bf Derived Parameters:} {\bf \em fs:} maximum fraction size of posit format given by $(ps-es-3)$ \\
        {\bf Initialize: } {\em k=0}
        
        
        \State $f0 \leftarrow \sim |P$  \Comment{\textcolor{blue}{\small set if 0}} \label{dec1}
        \State $fNaR \leftarrow P[ps-1] \mathbin{\&} ( \sim | (P[ps-2:0])$  \Comment{\textcolor{blue}{\small set if NaR}} \label{dec2}
        
        \State $s \leftarrow P[ps-1]$
        \If {($s = 1 $)} \label{dec3}
            \State $ P \leftarrow \sim P+1$ \Comment{\textcolor{blue}{\small 2's complement of P}} \label{dec4}
        \EndIf 
        \State $t \leftarrow P[ps-1:0] $ \label{dec5}
        \If {($P[ps-2] = 1$)}
            \State $t \leftarrow \sim t$ \Comment{\textcolor{blue}{\small 1's complement of $t$}}
        \EndIf
        \State  $ rc \leftarrow countZerosMSB(t)$ \label{dec6}
        
        \If {($P[ps-2] = 1$)} \label{dec7}
            \State $k \leftarrow rc-1$
        \Else
            \State $k \leftarrow (-ve)rc$
        \EndIf \label{dec8}
        \State $ P \leftarrow P \ll rc+2 $ 
        \State $e \leftarrow P[ps-1:ps-es] $ \label{dec9}
        \State $exp \leftarrow e+ k \ll es $ \label{dec10}
         \State $P \leftarrow P \ll es $ \label{dec11}
         \State $f \leftarrow 1 \mathbin\Vert P[ps-1:ps-fs] $ \Comment{\textcolor{blue}{\small append hidden bit}} \label{dec12}
         
 \\     \Return  {\em s, exp, f, f0, fNaR}
    \end{algorithmic} 
 \end{algorithm}

\subsection{Common Posit Decoder}
\label{sec:posit_decode}
This is the first block of our \posit unit and is responsible for extracting the sign, exponent,
and fraction of each posit-number, {\em P}, introduced at its inputs. This unit is also responsible for detecting if the inputs are 0 or NaR and set the appropriate flags. The functionality of this
unit is captured in Algorithm-\ref{algo:decode_posit}. 
 
Compared to \float, which requires detecting five different special values: subnormal, zero, qNan, sNan and infinity, \posit requires detecting only two special values. Unlike \float, the sign bit
is the only field which has a fixed position in a \posit number. This sign bit indicates if the
number is negative and if it is, a 2's complement of the number should be taken (lines \ref{dec3}-\ref{dec4}). To extract
the exponent and fraction, we count the number of regime bits (lines~\ref{dec5}-\ref{dec6}). If the regime
starts with a 0, then the exponent is treated as negative, else positive. Please note, that the sign
of the exponent in \float is derived through a bias. Lines~\ref{dec7}-\ref{dec10} 
implement Equations-\ref{eq:k_value} and \ref{eq:exp} to capture the final exponent value. Lastly, fraction bits are obtained by shifting out the exponent bits (lines \ref{dec11}-\ref{dec12}). Unlike \float, \posit do not have subnormal numbers, eliminating the need to handle subnormal exponent and fraction separately, thereby simplifying the \posit arithmetic.

\subsection{Common Posit Encoder}
\label{sec:posit_encode}

This is the final block of the design and caters to creating the final
posit representation from the results computed by various arithmetic compute
blocks. The capabilities of this block are captured in Algorithm-\ref{algo:encode_posit}.
To calculate the final exponent we first extract the \emph{e} bits and the value of 
\emph{k} (lines \ref{enc1}-\ref{enc2}). A negative exponent is represented with 
the regime bits beginning with a 1 (0 otherwise) (lines \ref{enc3}-\ref{enc4}). 
The sticky bit (\emph{sb}) is calculated as the Logical-OR of all the shifted-bits. 
Once the regime bits have been deduced, the exponent and fraction bits are appended
appropriately (lines \ref{enc7}-\ref{enc8}). Following this, the posit is rounded
to the nearest even (with tie-to-even) in lines \ref{enc9}-\ref{enc10}. If either of
the input flags, 0 or NaR, are set, then the final result is modified accordingly (lines \ref{enc11}-\ref{enc12}), 
else the rounded result is forwarded to the output.

The \posit scheme differs significantly from the \float scheme when it comes to rounding.
Only one rounding mode is supported in \posit as compared to five in \float. 
A \posit number neither overflows nor underflows. If the encoded number is {\em maxpos} (integer value of $2^{ps-1}-1$), it is not rounded up irrespective of round bit (\emph{rb}). Similarly, if the encoded number is {\em 0}, it is rounded up irrespective of the round bit.

 \begin{algorithm}[!t]
 \caption{Algorithm for \Posit Encoding}
 \label{algo:encode_posit}
 \begin{algorithmic}[1]
 \renewcommand{\algorithmicrequire}{\textbf{Input:}}
 \renewcommand{\algorithmicensure}{\textbf{Output:}}
 \Require {\bf \em rs:} sign for {\bf \em P}, {\bf \em rexp:} exponent for {\bf \em P}, {\bf \em rf:} fraction bits for {\bf \em P}, {\bf \em sb:} sb is set if any of previous shifted bits are 1, {\bf \em f0:} if {\bf \em P} is 0, {\bf \em fNaR:} if {\bf \em P} is NaR
 \Ensure  {\bf \em P}: final result in posit format with $ps$ bits\\
 {\bf Derived Parameters:} {\bf \em fes}: maximum exponent size of {\em ps} bit posit, given by $(\log_2(ps)+es+2)$ \\
 {\bf Initialize: } {\em P = 0, c=2}
 \State $ e \leftarrow rexp[es-1:0] $ 
 \label{enc1}
 \State $ k \leftarrow rexp[fes-1:es] $
 
 \label{enc2}
 \If{($ k \geq 0 $)} \label{enc3}
 \State $P \leftarrow \sim P $ \Comment{\textcolor{blue}{\small regime start with 1 for positive exp}}
 \State $c = 3$
 \Else
 \State $ P[0] \leftarrow 1 $
 \EndIf 
 \State $ k' \leftarrow abs(k) $ 
    \State $ P \leftarrow P \ll (ps+1-k`)$ \label{enc5} \Comment{\textcolor{blue}{\small shift P to finalize regime}} \label{enc6}
 \State $P[ps] \leftarrow 0 $ \Comment{\textcolor{blue}{\small add sign bit}} \label{enc4}
 

 \State $esft \leftarrow k' + c$ \label{enc7}
    \State $ ef \leftarrow e \mathbin \Vert rf $ \label{enc:es_check}
    \State $sb \leftarrow sb \mathbin{|} (\mathbin{|}(ef \ll  ps+1-esft))  $ \Comment{\textcolor{blue}{\small update sb}}
    \State $ef \leftarrow ef \gg esft$ \Comment{\textcolor{blue}{\small adjust e,f wrt regime}}
 \State $P \leftarrow P \mathbin{|} ef$ \label{enc8} \Comment{\textcolor{blue}{\small e and frac bits are added}} 
 \State $rb \leftarrow P[0]$, $gb \leftarrow P[1]$ \label{enc9}
 
 \State $ rb \leftarrow rb$ \& $gb$ $|$ $sb $ \Comment{\textcolor{blue}{\small set rb as per round-to-nearest}}
      \State $max \leftarrow \sim P[ps-1]) \& (\& P[ps-2:0] $ 
      \If{($max = 1$)} \Comment{\textcolor{blue}{\small handles overflow}}
        \State $rb \leftarrow 0 $ \Comment{\textcolor{blue}{\small do not round if P is maxpos}}
     \EndIf
     \If {($P = 0$)} \Comment{\textcolor{blue}{\small handles underflow}}
     \State $P \leftarrow 1 $ \Comment{\textcolor{blue}{\small round up if P is zero}}
     \EndIf
      \State $rb \leftarrow rb \oplus s $
     \If{($rs=1$)} \Comment{\textcolor{blue}{\small complement negative number}}
     \State $P \leftarrow \sim P $
     \EndIf
     \State $P \leftarrow P + rb $ \label{enc10}
     \If{($f0 = 1$)} \Comment{\textcolor{blue}{\small P is 0, if f0 is set}} \label{enc11}
        \State $P \leftarrow 0 $
     \EndIf
     \If{($fNaR = 1$)} \Comment{\textcolor{blue}{\small P is NaR if fNaR is set}}
        \State $P \leftarrow NaR $ \label{enc12}
     \EndIf 
     \\ \Return  {\em P}
     
 \end{algorithmic} 
 \end{algorithm}

The following subsections will define the different modules and the instruction they implement in the proposed \posit unit.

\subsection{Fused Multiply-Add (FMA)}
\label{sec:fma}

 \begin{algorithm}[!t]
 \caption{Algorithm for Fused Multiply-Add}
 \label{algo:fma}
 \begin{algorithmic}[1]
 \renewcommand{\algorithmicrequire}{\textbf{Input:}}
 \renewcommand{\algorithmicensure}{\textbf{Output:}}
 \Require { \bf \em s\textsubscript 1, exp\textsubscript 1, f\textsubscript 1, f0\textsubscript 1, fNaR\textsubscript 1:} components of operand\textsubscript 1 from decode stage, { \bf \em s\textsubscript 2, exp\textsubscript 2, f\textsubscript 2, f0\textsubscript 2, fNaR\textsubscript 2:} components of operand\textsubscript 2, { \bf \em s\textsubscript 3, exp\textsubscript 3, f\textsubscript 3, f0\textsubscript 3, fNaR\textsubscript 3:} components of operand\textsubscript 3, {\bf \em ng:} if negate operation, {\bf \em op:} if sub operation 
 \Ensure {\bf \em rs:} sign of result, {\bf \em rexp} final exponent of fma,  {\bf \em rf:} final fraction of fma, {\bf \em sb:} sticky bit, {\bf \em f0:} set if result is 0, {\bf \em fNaR:} set if result is NaR \\
 {\bf Derived Parameters:} {\bf \em ffs:} final fraction size for fma given by $(2 \times (ps-es-2))$ \\
 {\bf Initialize: } {\em } 

 \If{($(fNaR_1|fNaR_2|fNaR_3) = 1$)} \label{fma1}
    \State $fNaR \leftarrow 1$  \Comment{\textcolor{blue}{\small final result is NaR}}
 \EndIf
 \If{((($f0_1 | f0_2) \text{\&} f0_3) = 1$)}
    \State $f0 \leftarrow 1$ \Comment{\textcolor{blue}{\small final result is 0}}
 \EndIf \label{fma2}
 
 \State $ s_3 \leftarrow s_3 \oplus op \oplus ng $ \Comment{\textcolor{blue}{\small sub operation affects sign of op\textsubscript 3}} \label{fma3} 
 \State $ rs \leftarrow s_1 \oplus s_2 \oplus ng $
 \State $ rexp \leftarrow exp_1 + exp_2 $
 \State $ rf \leftarrow f_1 \times f_2 $
 \State $ rexp, rf \leftarrow chkMulOF(rf)$ \label{fma4}
 \If{($(exp_3 > rexp) || (exp_3 = rexp \text{\&} f_3 > rf )$ )} \label{fma5}
 \State $swap(rs,s_3)$, $swap(rexp,exp_3)$, $swap(rf,f_3)$
 \EndIf 
 \State $ ediff \leftarrow rexp - exp_3 $
 \State $ sb \leftarrow |(f_3 \ll (ffs-ediff))$ \Comment{\textcolor{blue}{\small shifted bits are ORed}}
 \State $ f_3 \leftarrow f_3 \gg ediff$
 
 \If{($rs = s_3$)}
 \State $ rf \leftarrow rf + f_3 $
 \State $rf, rexp \leftarrow chkAddOF(rf) $
 \Else
 \State $ rf \leftarrow rf - f_3 $
 \State $rf, rexp \leftarrow normalize(rf) $
 \EndIf \label{fma6}
 \\ \Return {$ rs, rexp, rf, sb, f0, fNaR $}
 \end{algorithmic} 
 \end{algorithm}
 
 The 'F' extension of RISC-V ISA specifies four different fused ops, FMADD.S, FMSUB.S, FNMSUB.S and FNMADD.S. These operations are carried out using Algorithm-\ref{algo:fma}. The {\em ng} input bit indicates a negate operation, while the {\em op} input bit indicates a subtract operation. We have configured this
 block to support not only fused operations but also simple operations like FADD.S, FSUB.S and FMUL.S, enabling maximum resource to be re-used across operations.
 
 Our FMA block checks whether either of the inputs are 0 or NaR. The corresponding circuitry in \float would require checking for 5 exceptional cases per operand. Fused operations in \float require checking the intermediate product exponent for underflow or overflow, which is absent in \posit. 
Additionally, \float also requires normalizing the product's fraction in the case of subnormal numbers, while \posit is burdened with only checking the overflow of the product's fraction.
 \Posit do not set any flag in the {\em pcsr} register in contrast to \float that can set NV, OF, UF and NX flags in {\em fcsr}.
 

\subsection{Division (FDIV)}
\label{sec:division}
This block implements the FDIV.S instruction as Algorithm-\ref{algo:div}. 
A divide by zero exception in \posit is captured by setting the DZ flag in \emph{pcsr}
(lines \ref{div1}-\ref{div2}). \float, on the other hand, has to account for setting all 5 flags for FDIV.S.
The sign and exponent of the result are quite simply calculated as per lines \ref{div3}-\ref{div4}. 
This block uses an iterative non-restoring division algorithm for computing the division of the fractions (line \ref{div5}). The number of cycles required for the operation is proportional to the size of the fractions. In each cycle, two iterations of non-restoring division are performed. The non-restoring algorithm returns the quotient and remainder, where the remainder part is used to calculate the sticky bit (line \ref{div6}). The fractional part is checked for normalization before passing these values to the encoding module (line \ref{div7}).

Similar to FMA, \float divison has to normalize subnormal numbers before fraction division while \posit incur no such hardware. Also, \float checks the division exponent for overflow and underflow while \posit do not. 

\begin{algorithm}[!t]
 \caption{Algorithm for Division}
 \label{algo:div}
 \begin{algorithmic}[1]
 \renewcommand{\algorithmicrequire}{\textbf{Input:}}
 \renewcommand{\algorithmicensure}{\textbf{Output:}}
 
 \Require { \bf \em s\textsubscript 1, exp\textsubscript 1, f\textsubscript 1, f0\textsubscript 1, fNaR\textsubscript 1:} components of operand\textsubscript 1 from decode stage, { \bf \em s\textsubscript 2, exp\textsubscript 2, f\textsubscript 2, f0\textsubscript 2, fNaR\textsubscript 2:} components of operand\textsubscript 2
 \Ensure {\bf \em rs:} sign of result, {\bf \em rexp} final exponent of division,  {\bf \em rf:} final fraction of division, {\bf \em sb:} sticky bit, {\bf \em f0:} set if result is 0, {\bf \em fNaR:} set if result is NaR, {\bf \em ex:} exception flags
 \textbf{Initialize: }
 
  \If{($(fNaR_1|fNaR_2) = 1$)} 
    \State $fNaR \leftarrow 1$
    \EndIf
    \If{($f0_2 = 0$)} \label{div1}
        \State $ex[3] \leftarrow 1$ \Comment{\textcolor{blue}{\small update DZ flag in {\em pcsr}}}  \label{div2}
    \EndIf
 \If{($f0_1 = 1$)}
    \State $f0 \leftarrow 1$
 \EndIf 
 
 \State $ rs \leftarrow s_1 \oplus s_2 $ \label{div3}
 \State $ rexp \leftarrow exp_1 - exp_2 $ \label{div4}
 
 
 

 \State $ rf, rem \leftarrow nonRstrDiv(f_1, f_2)$  \Comment{\textcolor{blue}{\small multi-cycle algo}} \label{div5}
 \State $ sb \leftarrow |rem $ \label{div6}
 \State $ rf, rexp \leftarrow normalize(rf)$  \label{div7}
 \\ \Return {$ rs, rexp, rf, sb, f0, fNaR, ex $}
 \end{algorithmic} 
 \end{algorithm}

\subsection{Square-Root (FSQRT)}
\label{sec:square-root}

Algorithm-\ref{algo:sqrt} captures the implementation details of the square-root operation (FSQRT.S). 
If the input is an NaR  or negative input the algorithm returns NaR (lines \ref{sqrt1}-\ref{sqrt2}). 
The exponent of the result is obtained by dividing the input exponent by two (line \ref{sqrt3}). 
If the input exponent is odd, the fraction is left-shifted by one (lines \ref{sqrt4}-\ref{sqrt5}).
Similar to division, we have implemented an iterative non-restoring square root algorithm (line \ref{sqrt6}), 
where each cycle performs a single iteration. Square root operation also does not set any flags in {\em pcsr}, while \float set the inexact flag. \float algorithms are forced to normalize any subnormal inputs before finding the root while \posits are not.
 
\begin{algorithm}[!t]
 \caption{Algorithm for Square-Root}
 \label{algo:sqrt}
 \begin{algorithmic}[1]
 \renewcommand{\algorithmicrequire}{\textbf{Input:}}
 \renewcommand{\algorithmicensure}{\textbf{Output:}}
 
 \Require { \bf \em s\textsubscript 1, exp\textsubscript 1, f\textsubscript 1, f0\textsubscript 1, fNaR\textsubscript 1:} components of operand\textsubscript 1 from decode stage
 \Ensure {\bf \em rs:} sign of result, {\bf \em rexp} final exponent of sqrt,  {\bf \em rf:} final fraction of sqrt, {\bf \em sb:} sticky bit, {\bf \em f0:} set if result is 0, {\bf \em fNaR:} set if result is NaR
 
 \If{($(fNaR_1|s_1) = 1$)} \Comment{\textcolor{blue}{\small return NaR for -ve input}}\label{sqrt1}
    \State $fNaR \leftarrow 1$ \label{sqrt2}
 \EndIf
 \If{($f0_1 = 1$)}
    \State $f0 \leftarrow 1$
 \EndIf
 \State $ rexp \leftarrow exp_1 \gg 1 $ \label{sqrt3}
 \If{($exp_1[0]==1$)} \label{sqrt4}
 \State $ f_1 \leftarrow f_1 \ll 1 $ \Comment{\textcolor{blue}{\small shift left for odd exponent}}
 \EndIf \label{sqrt5}
    
 

 \State $ rf, rem \leftarrow nonRstrSqrt(f_1) $ \Comment{\textcolor{blue}{\small multi-cycle algo}} \label{sqrt6}
 \State $ sb \leftarrow |rem $
 \\ \Return {$ 0, rexp, rf, sb, f0, fNaR $}

 \end{algorithmic} 
 \end{algorithm}

\subsection{Integer to Posit Conversion}
\label{sec:int_to_posit}

FCVT.S.W/FCVT.S.WU instructions are used to convert a signed/unsigned value to \posit respectively. The conversion steps are highlighted in Algorithm-\ref{algo:int_to_posit}. The {\em u} input bit indicates if the input is unsigned, in which case the negative sign is cleared (line \ref{itop1}). To get the actual exponent of \posit, we set the maximum exponent according to the {\em ps} value. Then we count the number of leading zeros and subtract this count from the maximum exponent. The integer value is shifted left to get the fractional part of \posit (line \ref{itop2}-\ref{itop3}). Integer to floating-point conversion operation for \posit do not set any exceptional flag while \float conversion involves setting of the inexact flag. In addition to the above, \float also requires check for special values of {\em n} bit signed integers (0 and $2^{n-1}$).

 \begin{algorithm}[!t]
 \caption{Algorithm for Integer to Posit Conversion}
 \label{algo:int_to_posit}
 \begin{algorithmic}[1]
 \renewcommand{\algorithmicrequire}{\textbf{Input:}}
 \renewcommand{\algorithmicensure}{\textbf{Output:}}
 
 \Require { \bf \em I:} integer of {\em ps} bits, {\bf \em u:} is set if {\bf \em I} is unsigned
 \Ensure {\bf \em rs:} sign of result, {\bf \em rexp:} final exponent of result,  {\bf \em rf:} final fraction of result
 
 \State $ rs \leftarrow I[ps-1] $
 \State $ rs \leftarrow rs $ \& $\sim u $ \Comment{\textcolor{blue}{\small do not consider sign if unsigned int}} \label{itop1}
 \If{($rs=1$)}
    \State $I \leftarrow \sim I + 1$ \Comment{\textcolor{blue}{\small 2's complement}}
 \EndIf
 
 \State $ z \leftarrow countZeroMSB(I) $ \label{itop2}
 \State $ I \leftarrow I \ll z $
 \State $ rexp \leftarrow ps - 1 - z$
 \State $ rf \leftarrow I[ps-2:0] $ \label{itop3}
 \\ \Return {$ rs, rexp, rf $}
 \end{algorithmic} 
 \end{algorithm}

\subsection{Posit to Integer Conversion}
\label{sec:posit_to_int}

The 'F' extension provides FCVT.W.S/FCVT.WU.S instructions to convert \posit to signed/unsigned integers respectively. These instructions are implemented using Algorithm-\ref{algo:posit_to_int}. The {\em u} input bit indicates if the result should be unsigned. The extended \posit fraction is left-shifted
by the exponent (lines \ref{ptoi1}-\ref{ptoi2}). The final integer is decided based on the {\em u} and {\em exp} inputs (lines  \ref{ptoi3}-\ref{ptoi4}). For FCVT.W.S and FCVT.WU.S instructions, we propose to support an additional rounding mode (round-to-zero mode) along with the default \posit rounding mode (motivation of this is discussed in Section-\ref{sec:image_proc}). Thus, when the rounding mode ({\em rm})  is round-to-zero, {\em rb} is set to 0 (line \ref{ptoi5}). \float performs several checks for setting the invalid and inexact flags during conversion to integer while \posit does not set any flags as a consequence of these
instructions.


 \begin{algorithm}[!t]
 \caption{Algorithm for Posit to Integer Conversion}
 \label{algo:posit_to_int}
 \begin{algorithmic}[1]
 \renewcommand{\algorithmicrequire}{\textbf{Input:}}
 \renewcommand{\algorithmicensure}{\textbf{Output:}}
 \Require { \bf \em s\textsubscript 1, exp\textsubscript 1, f\textsubscript 1:} components of operand\textsubscript 1 from decode stage, {\bf \em u:} is set if result is unsigned, {\bf \em rm}: rounding mode
 \Ensure {\bf \em I:}  integer value corresponding to posit \\
 {\bf Derived Parameter:} {\bf \em fs:} maximum fraction size of posit format given by $(ps-es-3)$
 
 \State $ f_1 \leftarrow 0_1 0_2 ... 0_{ps} \mathbin\Vert f_1  $ \Comment{\textcolor{blue}{\small zero extend f\textsubscript 1 to ps+fs bits}} \label{ptoi1}
 \State $ f_1 \leftarrow f_1 \ll exp_1 $ \label{ptoi2}
 \If{($u=0$)} \label{ptoi3}
    \If{($exp_1 < ps-1$)}
        \State $ I \leftarrow f_1[ps+fs-1:fs] $
    \Else
        \State $ I \leftarrow 2^{ps-1} - 1 $
    \EndIf
 \Else 
    \If{($exp_1 < ps$)}
        \State $ I \leftarrow f_1[ps+fs-1:fs] $
    \Else
        \State $ I \leftarrow 2^{ps} - 1 $
    \EndIf
 \EndIf \label{ptoi4}
 \State $ rb \leftarrow f_1[fs-1] $
 \If{($ rm = 1 $)} \label{ptoi5} \Comment{\textcolor{blue}{\small check for round-to-zero}}
    \State $ rb = 0$
 \EndIf
 \State $ I \leftarrow round(I, s_1, rb) $
 \\ \Return {\em I}
 \end{algorithmic} 
 \end{algorithm}

\subsection{Posit Comparison}
\label{sec:posit-comparison}
FMIN.S, FMAX.S, FEQ.S, FLT.S and FLE.S instructions are defined in the 'F' extension for comparison between floating-point numbers. The \posit representation resembles the 2's complement representation of an integer. Hence, the comparison between two \posits is exactly similar to comparing the integer value of the bit representation. There are no exceptional cases in \posit comparison operation. This is one of the simplifications in \posit arithmetic compared to \float arithmetic. Additionally,  We have utilized integer comparison logic for \posit comparison; hence, the need for a comparator in FPU is eliminated.
Comparison in \float is quite burdensome as compared to \posit. \float requires checking of different 
exceptional cases such as +0 and -0 since they have different representations but are treated as equal.
Similarly, two NaNs having similar bit representations are treated as different. \float shall set the NV flag for comparison operations (whenever applicable) while \posit does not set any flags. 
 
\subsection{Posit Sign Injection}
\label{sec:sign_inject}

RISC-V ISA defines FSGNJ.S, FSGNJN.S, FSGNJX.S instructions for floating-point sign injection. 
FSGNJ.S is used to move values between registers, FSGNJN.S is used to negate a floating-point number, and FSGNJX.S is used to get the absolute value of floats. The negation operation for a \float number is just flipping of sign bit, but in case of \posit number, 2's complement is taken to negate a number or to calculate the absolute value of a negative number. Both \posit and \float do not set any exception flags for this operation.

\subsection{Posit Classification}
\label{sec:classify}

FCLASS.S instruction has been defined to determine the category of a floating-point number. It is much required for \float but not so much with \posit. \float have different categories which are $\pm$0, $\pm$infinity, $\pm$subnormals, $\pm$normals, sNaN and qNaN. Hence \float need to check for all those values. \Posit only need to check for 0, NaR, +ve and -ve numbers, hence the logic is much simpler. Classify instruction also does not set any flags for either \posit or \float. 

\subsection{Dynamic Switching}
\label{sec:dynamic_switching}

Several scientific applications like weather forecasting, automotive design and safety etc. demand high precision floating-point calculations. On the other hand, applications in the deep learning domain require large dynamic ranges. One would require two separate designs to cater to both the demands. In regards to this, we propose a single \posit FPU which can fulfil the requirement of high precision and high dynamic range applications within the same design. We enhance the proposed hardware unit of 32-bit \posit to support two different \emph{es} values (\emph{ es=2 and es=3}). The \posit unit can, thus, switch across various {\em es} values at run-time by manipulating the \emph{es-mode} field in the \emph{pcsr} register. We call this capability as \emph{Dynamic Switching}.

For supporting dynamic switching, we would also need instruction support to convert \posit numbers encoded with one \emph{es} value to a \posit number encoded with a different 
\emph{es} value. In view of this, we introduce a new single-operand instruction: \emph{FCVT.ES} formatted 
as shown in Table-\ref{tab:dynamic_switching}. The \emph{from-es} field indicates the 5-bit
\emph{es} value with which the current register rs1 has been encoded, while the 
\emph{to-es} field indicates the target \emph{es} value to which the number should be 
encoded with. While switching from one {\em es} value to another, if the number is not exactly representable in the target {\em es} domain, \posit rounding logic ensures that the number is rounded correctly.
One should note that this instruction does not use the \emph{es-mode} present in the \emph{pcsr} register.

\begin{table}[!t]                                                                                  
{\footnotesize                                                                                      
\begin{center}                                                                                      
\begin{tabular}{E@{}E@{}cccccc}
\instbitrange{31}{25} &
\instbitrange{24}{20} &
\instbitrange{19}{15} &
\instbitrange{14}{12} &
\instbitrange{11}{7} &
\instbitrange{6}{0} \\
\hline      
\multicolumn{1}{|c|}{1111100} &
\multicolumn{1}{c|}{to-es} &
\multicolumn{1}{c|}{from-es} &
\multicolumn{1}{c|}{000} &
\multicolumn{1}{c|}{rs1/rd} &
\multicolumn{1}{c|}{opcode} \\
\hline

\end{tabular}
\end{center}
}                                                                                                   
\vspace{-0.1in}                                                                                     
\caption{Instruction format for switching operands across \emph{es} values.}                                               
\label{tab:dynamic_switching}    
\end{table}


To support two different {\em es} values we have made few modifications in the parameterized design presented in Section-\ref{sec:posit_arith}. The modifications are mentioned below. 
For a compile-time defined {\em es} values, these sizes are already mentioned in Algorithm-\ref{algo:decode_posit} and \ref{algo:encode_posit} as derived parameters. In case of run-time defined {\em es} value, the exponent size is determined by the largest \emph{es} value (\emph{es=3}) and the fraction size by smallest \emph{es} value(\emph{es=2}). Along with this, few changes are incorporated in encode and decode modules to handle two \emph{es} values.

For the \posit decode module, Equation-\ref{eq:es_check1} adjusts {\em e} bits for {\em es=2} after line \ref{dec9} of Algorithm-\ref{algo:decode_posit}.
\begin{equation}
\label{eq:es_check1}
    e = e \gg 1 
\end{equation}

For the \posit encode module in Algorithm-\ref{algo:encode_posit}, Equation-\ref{eq:es_check2} modifies {\em e} bits after line \ref{enc1} for {\em es = 2}. Where as Equation-\ref{eq:es_check3} adjusts {\em k} value after line \ref{enc2} for {\em es = 3}. Lastly, Equation-\ref{eq:es_check4} after line \ref{enc:es_check} shifts the concatenated {\em ef} valuer for {\em es = 2}.
\begin{equation}
\label{eq:es_check2}
    e[2] = 0 
\end{equation}
\vspace{-15pt}
\begin{equation}
\label{eq:es_check3}
    k = k \gg 1 
\end{equation}
\vspace{-15pt}
\begin{equation}
\label{eq:es_check4}
    ef =  ef \ll 1 
\end{equation}

 To switch from one {\em es} value to another, we first decode the \posit number as per the {\em from-es} value and then encode the sign, exponent and mantissa as per the {\em to-es} value. The encode and decode modules are already available in the \posit FPU, and the minimal changes mentioned above allows them to support two {\em es} values. Thus, the overheads for dynamic switching are minimal and discussed in Section-\ref{sec:hw-results}.
 

\section{Integration with Core}
\label{sec:core-integration}

\begin{figure}[t]
    \centering
    \includegraphics[width=\linewidth]{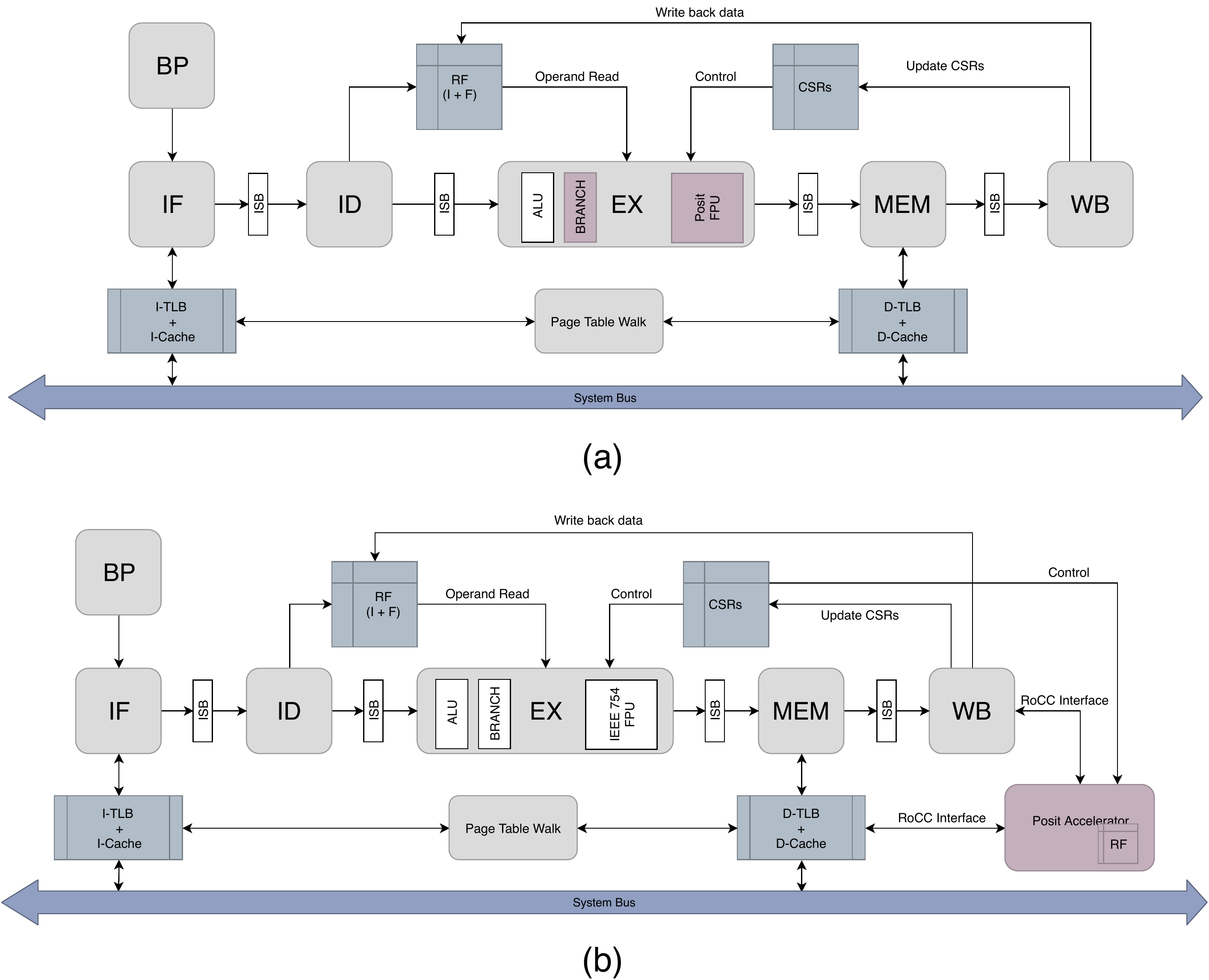}
    \caption{Micro-architecture of the \cclass core with \posit FPU integrated. (a) \posit FPU integrated as a tightly-coupled execution unit. (b) \posit FPU integrated as an accelerator through RoCC}
    \label{fig:cclass-posit}
\end{figure}

This section describes how the proposed \posit FPU can be integrated with a standard
RISC-V core. We have chosen the SHAKTI \cclass core~\cite{shakti} as our baseline
core for integration. \cclass is amongst one of the most configurable Linux-capable, 
in-order, RISC-V open-source core available. The \cclass core is designed using Bluespec-System-Verilog 
(BSV)~\cite{bluespec} and is a basic 5-stage in-order
core which includes: a branch predictor, blocking instruction and data caches, fully-associative
TLBs for instruction and data, a HW page-table-walk unit and AXI-4 compliant system bus interface.
Figure-\ref{fig:cclass-posit} shows a high-level micro-architecture of the core. 

For the purpose of this work, we have 
configured the core to support the RV32IMAFC extensions of the ISA (i.e. it supports, 32-bit Integer (I),
multiplication/division (M), atomic (A), single-precision floating (F) and compressed (C) ISA extensions).
The \posit FPU has been instantiated with parameter settings: \emph{ps=32} and \emph{es=2}. The next subsections elaborate further on how the proposed \posit FPU can be integrated as either a 
tightly-coupled execution unit or as an accelerator through RoCC interface.

\subsection{Integration as a tightly-coupled execution unit}
\label{sec:tightly-coupled}
The default \cclass core includes a single-precision \float compliant floating point unit which has the following BSV interface definition.  \\

\begin{Verbatim}[commandchars=\\\{\},frame=single,fontsize=\footnotesize,samepage=true]
interface Ifc_fpu;
  method Action _start(
    Bit#(32) rs1, Bit#(32) rs2, Bit#(32) rs3,
    Bit#(4) opcode, Bit#(7) funct7, Bit#(3) funct3,
    Bit#(2) imm, Bit#(3) csr);
  method ActionValue#(Bit#(32)) get_rd;
  method ActionValue#(Bit#(5)) get_fflags;
endinterface
\end{Verbatim}

Since our \posit FPU implements the same interface (except for the CSR ({\em es-mode} in case of \posit) field being 5-bits in \posit to include the \emph{es} value), we are
able to effortlessly replace the \float unit with our \posit unit.

As mentioned in Section-\ref{sec:posit-comparison}, the comparison operations for \posit 
can re-use the integer comparison hardware blocks. In the \cclass, the branch unit in the execution
stage performs the comparison of signed/unsigned integers. We modify this block to receive inputs
for \posit comparison instructions as well. 

Figure-\ref{fig:cclass-posit}(a) shows the micro-architecture
of the \posit FPU integrated with the \cclass core as a tightly-coupled execution unit.
The flag updates in the \emph{pcsr} happen similar to how the \float unit updates the flags, i.e.
in the write-back stage.

\subsection{Integration as an accelerator}
\label{sec:accelerator}
Figure-\ref{fig:cclass-posit}(b) shows how the proposed \posit FPU is integrated with the \cclass 
core as an accelerator. The \posit FPU is connected to the core at the write-back stage through the
RoCC interface. The \posit FPU also is given access to the data-cache through the RoCC interface
to carry out memory-related operations. A major difference here as compared to the previous approach
of integration is that the posit register file is maintained within the accelerator rather than in the core. The register-file within the core would include the integer and \float register files.

As explained in Section-\ref{sec:custom-isa}, we leverage the \emph{major-custom} opcodes
of the RISC-V ISA to facilitate a \posit based arithmetic accelerator for integration with a RISC-V 
core. When an instruction containing any of custom opcodes is detected in the decode stage, 
the \emph{xs1/xs2} fields are checked to see if any of integer registers are required for the current
custom instruction. The execute and memory stage simply bypass this instruction without any changes. 
The write-back stage off-loads this instruction (along with the integer operands if any) to the \posit accelerator and waits for an execution-complete response from the accelerator. Depending on the value of the \emph{xd} field in the custom instruction, the write-back updates the relevant integer registers with the response received from the accelerator. 

With this approach, we are able to empower the existence of \float and \posit FPU in the same design
and enable the \posit community to develop more sophisticated and powerful \posit compute blocks without
having to touch the core-pipeline.

\subsection{Verification of Posit FPU}
We have used SoftPosit~\cite{softposit} library for verifying our \posit implementation. Random inputs were generated and the corresponding outputs of each operation(\emph{es=2}) were found to be in agreement with the result of the soft-posit library. Along with the random tests, we have also verified our designs for special cases. Apart from the two exceptional values(0 and \emph{NaR}) defined in \posit representation, we have also verified our design for the smallest and largest positive as well as negative numbers that can be represented in the \posit system (for both \emph{es=2} and \emph{es=3}).

\section{Software Support for Posit}
\label{sec:sw_support}
Since \Posit has only recently been introduced, to the best of our knowledge compiler support for \posits is not available for RISC-V. In light of this, some recent works such as ~\cite{date, calligo, vividsparks} suggest converting \float to \posit in hardware, thereby allowing compatibility with existing tool-chain. However,  we have observed that such conversion schemes can not leverage the benefits of \posit format. The dynamic range and precision that can be expressed are still limited by a 32-bit \float. For, eg. we cannot represent the number 3.0E+40 in 32-bit \float representation whereas it can be represented in 32-bit \posit(es=3) as 3.000865123284026E+40. The dynamic range of 32-bit \posit 
(for es=3 being 2.0E-75 to 5.0E+74) is greater than 32-bit \float(7.0E-46 to 3.0E+38). Similarly, 15.996093809604645 is represented in 32-bit \float as 15.99609375 because \float fraction have a precision of 24 bits, whereas it is exactly represented in 32-bit \posit(es=2) because \posit fraction have a \emph{max-precision} of 28 bits. 

\par Another approach would be to store the data as a double-precision \float and convert it to 32-bit \posit in hardware. While this approach represents a short-time solution, it defeats the original purpose of having \posits. The initial motivation to switch to \posit was the memory wall problem. It is well known that once we reach the memory wall, the program execution time will depend almost entirely on the speed at which RAM can send data to the CPU. The intention of using \posit was to replace 64-bit data with 32-bit data and thereby reduce the bandwidth requirement by half. 

\par From the above arguments, we conclude that a complete software stack support is the only way forward to leverage the true potential that \posit offers. While the up-streamed version of GCC supports the latest RISC-V ISA, we acknowledge the fact that the effort to provide
\posit support based on the proposals made in Section-\ref{sec:isa-extension} is significant and beyond the scope of this work. Moreover, providing custom instruction support (as mentioned in Section-\ref{sec:custom-isa}) to GCC will be even more challenging. In consideration of this situation, we adopt the following approach to modify basic C/C++ applications to facilitate porting of these applications on our \posit enable \cclass core. Here, we leverage the 'F' extension provided by RISC-V as is and integrate the \posit unit as a tightly-coupled execution
unit in the \cclass core as specified in Section-\ref{sec:tightly-coupled}.

Whenever a float variable is initialized in a program, the compiler assigns it the value according to \float representation. For our purpose, we require this value to be represented as a \posit equivalent. 
We achieve this by assigning the hexadecimal value of \posit representation (computed manually) to an 
integer variable and using \emph{memcpy} to transfer the contents of the integer variable to float 
variable. The below code snippet is an example of floating-point addition. \\
\begin{Verbatim}[commandchars=\\\{\},frame=single,fontsize=\footnotesize ,samepage=true]
float f1pt5 = 1.5;
float f1pt2 = 1.2;
float a = f1pt5 + f1pt2;
\end{Verbatim}
In the above example, the float variables f1pt5 and f1pt2 have been assigned values as per the \float format. We change this to equivalent \posit format, as shown in the code snippet below: \\
\begin{Verbatim}[commandchars=\\\{\},frame=single,fontsize=\footnotesize,samepage=true]
float f1pt5 = 1.5;
float f1pt2 = 1.2;

\textcolor{blue}{int i1pt5 = 0x44000000;}
\textcolor{blue}{memcpy(&f1pt5, &i1pt5, 4);}
\textcolor{blue}{int i1pt2 = 0x4199999A;}
\textcolor{blue}{memcpy(&f1pt2, &i1pt2, 4);}

float a = f1pt5 + f1pt2;
\end{Verbatim}
Similarly, when a float constant is used in the program, we need to replace it with the \posit equivalent. In the below code MAX is defined as a float constant. The compiler replaces MAX by its value during compilation. \\
\begin{Verbatim}[commandchars=\\\{\},frame=single,fontsize=\footnotesize,samepage=true]
#define MAX 1.5;
float f1pt2 = 1.2;
if(f1pt2 < MAX) ...
\end{Verbatim}
In the below snippet, we have used a float variable instead of constant. Now we can assign the \posit equivalent of float constant as mentioned above.\\
\begin{Verbatim}[commandchars=\\\{\},frame=single,fontsize=\footnotesize,samepage=true]
float f1pt2 = 1.2;

\textcolor{blue}{float fmax;}
\textcolor{blue}{int imax = 0x44000000;}
\textcolor{blue}{memcpy(&fmax, &imax, 4);}
\textcolor{blue}{int i1pt2 = 0x4199999A;}
\textcolor{blue}{memcpy(&f1pt2, &i1pt2, 4);}

\textcolor{blue}{if(f1pt2 < fmax)} ...
\end{Verbatim}

\par We acknowledge the fact that making compilers compatible with \posits can take up some time and the computing stack will not be complete for \posit until such support is available. However, the above-mentioned solution serves our purpose of running applications on \posit based RISC-V core.

\section{Applications}
\label{sec:applications}

In this section, we present details of porting several applications on our \posit enabled \cclass core. Each of these applications has been modified as per
Section-\ref{sec:sw_support} and executed on the \cclass core with \posit FPU integrated as a tightly-coupled execution unit. Additionally, we also provide
results and insights of running the same applications with \float on the default
\cclass core.


\subsection{Image Processing}
\label{sec:image_proc}



\begin{figure*}[!t]
\centering
  \begin{subfigure}{0.15\textwidth} \centering
     \includegraphics[width=2.2cm]{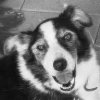}
     \caption{JPEG compress}\label{ip1}
  \end{subfigure}
  \begin{subfigure}{0.15\textwidth} \centering
     \includegraphics[width=2.2cm]{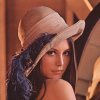}
     \caption{Original}\label{ip2}
  \end{subfigure}
  \begin{subfigure}{0.15\textwidth} \centering
     \includegraphics[width=2.2cm]{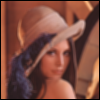}
     \caption{Blurred}\label{ip3}
  \end{subfigure}
  \begin{subfigure}{0.15\textwidth} \centering
     \includegraphics[width=2.2cm]{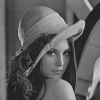}
     \caption{Gray}\label{ip4}
  \end{subfigure}
  \begin{subfigure}{0.15\textwidth} \centering
     \includegraphics[width=2.2cm]{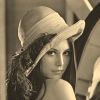}
     \caption{Sepia}\label{ip5}
  \end{subfigure}
  \begin{subfigure}{0.15\textwidth} \centering
     \includegraphics[width=2.2cm]{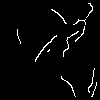}
     \caption{Edge detect}\label{ip6}
  \end{subfigure}
\caption{Images used in image processing applications} \label{image_proc}
\end{figure*}

We performed image processing tasks like JPEG compression, image filtering and edge detection with \posit and \float FPUs on the images given in Figure-\ref{image_proc}. In all the applications, the image inputs were represented as an array of integer values. We compressed three different variants of the image given in Figure-\ref{ip1} using \float and \posits. The original and compressed image sizes (in KBs) are given in Table-\ref{tab:jpeg}. We observed that \posit, with the default rounding mode, produced larger compressed images as compared to those obtained using \float. The culprit, based on our analysis, seems to the rounding mode used while converting \posit to integer. However, we observed that if conversion operation supported the RTZ (round-to-zero) rounding mode, the compression quality matched with those of \float.
This observation and study form the basis of our proposal to maintain two rounding modes for the \posit-to-integer instruction in Section-\ref{sec:posit_to_int}.

\begin{scriptsize}
\begin{table}[!t]
    \small
    \begin{tabular}{|C{0.6cm}|C{1.5cm}|C{1.4cm}|C{1.4cm}|C{1.5cm}|}
      \hline
      \textbf{\#} & \textbf{Original} & \textbf{Posit (RNE)} & \textbf{Posit (RTZ)} & \textbf{IEEE 754} \\ \hline \hline
      1 & 3.7KB & 1.9KB & 1.5KB & 1.5KB \\ \hline
      2 & 4.3KB & 2.3KB & 2.0KB & 2.0KB \\ \hline
      3 & 5.6KB & 3.0KB & 2.5KB & 2.5KB \\ \hline
    \end{tabular}
    \caption{Original and compressed image size in kBs after JPEG compression for 32-bit \posits and \float }
    \label{tab:jpeg}
\end{table}
\end{scriptsize}


The image filtering application took a bitmap image (Figure-\ref{ip2}) as input and applied filters like blur, converting to grayscale and converting to sepia. The respective images are given in Figure-\ref{ip3}, Figure-\ref{ip4}, Figure-\ref{ip5}. With RTZ rounding mode in \posit, we obtained similar results for 32-bit \float and \posits for this application. Similarly canny edge detection technique, which was used to detect edges in Portable Gray Map (PGM) images, was able to detect similar edges (Figure-\ref{ip6}) for \float and \posits with RTZ rounding.

\subsection{Trigonometric and Exponential Series}
\label{sec:trig_series}

Applications using trigonometric equations form an excellent platform to demonstrate how \posit outperform \float in terms of accuracy. We calculated sine, cosine, and exponential values for 32-bit \float and \posit(\emph{es=2}) using power series. The input for sine and cosine spans across values between 0-359  degrees. In case of e\textsuperscript{x} input takes values between 0-11. For our metric of comparison,
we have chosen the mean of the percentage error with respect to a double-precision \float result.
We further calculate the confidence interval~\cite{confidence_interval} with confidence of 95\%
to estimate the range of mean percentage error. These results are tabulated in Table-\ref{tab:percent_error}. We observed that the mean percentage error for \posit is 7x less in case of sin(x), 5x less in case of cos(x) and 7x less in case of e\textsuperscript{x} as compared to \float. Also, the confidence interval is non-overlapping in all cases. Thus, we conclude from Table-\ref{tab:percent_error} that \posit stores more information than \float for the same bit width; hence providing better accuracy.

\begin{scriptsize}
\begin{table}[!t]
    \small
    \begin{tabular}{|C{1.4cm}|C{1.6cm}|C{1.6cm}|C{2cm}|}
    \hline
      \textbf{Function} & \textbf{Metric} &\textbf{Posit} & \textbf{IEEE 754 Float}\\ \hline \hline
      \multirow{2}{*}{sin(x)} 
       & mean & 3.50E-05 & 2.32E-04 \\ \cline{2-4}
       & confidence interval & (1.20, 5.80) E-05 & (7.63, 38.8) E-05 \\
      \hline
      \multirow{2}{*}{cos(x)} 
       & mean & 2.18E-05 & 1.18E-04 \\ \cline{2-4}
       & confidence interval & (1.45, 2.86) E-05 & (8.56, 15.1) E-05 \\
      \hline
      \multirow{2}{*}{e\textsuperscript{x}} 
       & mean & 9.66E-07 & 6.30E-06 \\ \cline{2-4}
       & confidence interval & (2.55, 16.8) E-07 & (2.85, 9.76) E-06 \\ 
      \hline
    \end{tabular}
    \caption{Mean and confidence interval of percentage error for 32-bit \posit and \float for trigonometric and exponential series}
    \label{tab:percent_error}
\end{table}
\end{scriptsize}

\subsection{Fast Fourier Transform (FFT)}
\label{sec:fft}

\begin{scriptsize}
\begin{table}[!t]
    \small
    \begin{tabular}{|C{1.8cm}|C{1.4cm}|C{1.6cm}|C{1.8cm}|}
    \hline
      \textbf{Component} & \textbf{Metric} &\textbf{Posit} & \textbf{IEEE 754 Float}\\ \hline \hline
      \multirow{2}{*}{magnitude} 
       & mean & 2.10E-05 & 2.56E-04 \\ \cline{2-4}
       & confidence interval & (8.82, 33.1) E-06 & (6.34, 44.8) E-05 \\ 
      \hline
      \multirow{2}{*}{angle} 
       & mean & 5.02E-06 & 4.95E-05 \\ \cline{2-4}
       & confidence interval & (4.16, 5.87) E-06 & (3.82, 6.08) E-05 \\ 
      \hline
    \end{tabular}
    \caption{Mean and confidence interval of percentage error for 32-bit \posit and \float for Fast Fourier Transform}
    \label{tab:fft}
\end{table}
\end{scriptsize}

Applications like designing and using antennas, image processing and filters, data processing and analysis, etc. use FFT for different purposes. We calculated FFT for a complex input vector with the real component as cosine values of numbers between 0-127 and imaginary component as sine values. The resultant complex vector was converted to polar form. For this vector, we calculated the percentage error of 32-bit \float and \posit(\emph{es=2}) results for magnitude and angle separately, with respect to double-precision \float. The values obtained in Table-\ref{tab:fft} shows that the mean percentage error for \posit is 12x less in case of magnitude and 10x less in case of angle with respect to \float.

\subsection{K-means Clustering}
\label{sec:kmeans}

K-means algorithm is widely used in data science. We clustered some well-known data sets~\cite{iris ,wisconsin, ecoli} using the k-means algorithm. After performing clustering, we calculated different cluster quality metrics using~\cite{scikit-learn}. Values of all the metrics lie between 0 and 1, and a score near to 1 indicates better clustering. These metrics require the true and predicted label of data to measure the cluster quality. The true labels are present in the data set and predicted labels are generated by 32-bit \float and \posit. We observed that 32-bit \float and \posit gave similar clusters. To prove the superiority of \posit over 32-bit \float, we generated a data set consisting of 100 instances of 1000 random points in two-dimension.  Here, the true labels are generated using double precision \float and predicted labels are the ones generated by 32-bit \float and \posit. 

\subsubsection{max-precision mode}
\label{sec:kmeans:max-precision}

32-bit \posit \emph{(es=2)} provides higher precision than 32-bit \float. We compared \posit and \float cluster using the cluster quality metrics to get the numbers of cases where \posit is similar as or outperform \float. Table-\ref{tab:kmeans_prec} shows the comparison of \posit and \float result. On average, \posit provide similar or better results than \float in 73\% cases.

\begin{scriptsize}
\begin{table}[!t]
    \small
    \begin{tabular}{|C{0.7cm}|C{1.8cm}|C{1.8cm}|C{2.8cm}|}
      \hline
      \textbf{k} & \textbf{Posit Passed} & \textbf{IEEE 754 Float Passed} & \textbf{No. of instances where \posit outperforms}\\ \hline \hline
      2 & 100 & 100 & 100 \\ \hline
      3 & 100 & 100 & 81 \\ \hline
      4 & 100 & 100 & 66 \\ \hline
      5 & 100 & 100 & 69 \\ \hline
      6 & 100 & 100 & 57 \\ \hline
      7 & 100 & 100 & 66 \\ \hline
    \end{tabular}
    \caption{Comparison of cluster quality for 32-bit \float and \posit in max-precision mode for synthetic data set with different k Values, where k indicates the number of clusters}
    \label{tab:kmeans_prec}
\end{table}
\end{scriptsize}

\subsubsection{max-dynamic range mode}
\label{sec:kmeans:max-dynamic-range}

\begin{scriptsize}
\begin{table}[!t]
    \small
    \begin{tabular}{|C{0.7cm}|C{1.8cm}|C{1.8cm}|C{2.8cm}|}
      \hline
      \textbf{k} & \textbf{Posit Passed} & \textbf{\float Passed} & \textbf{No. of instances where \posit outperforms}\\ \hline \hline
      2 & 100 & 100 & 100 \\ \hline
      3 & 100 & 78 & 36 \\ \hline
      4 & 100 & 46 & 23 \\ \hline
      5 & 100 & 29 & 16 \\ \hline
      6 & 100 & 28 & 15 \\ \hline
      7 & 100 & 0 & 0 \\ \hline
    \end{tabular}
    \caption{Comparison of cluster quality for 32-bit \float and \posit in max-dynamic range mode for synthetic data set with different k Values, where k indicates the number of clusters}
    \label{tab:kmeans_range}
\end{table}
\end{scriptsize}

The dynamic range of 32-bit \posit\emph{(es=3)} is greater than 32-bit \float. Hence, \posit can handle much larger values compared to \float. We multiplied the random input with a large number and clustered them using \float and \posits. Table-\ref{tab:kmeans_range} shows the comparison of \posit and \float result. \float did not pass all the cases due to overflow during the calculation of large values, whereas \posit passed all the cases as it can handle a larger dynamic range. Out of all the cases in which \float passes, 32-bit \posit provide similar or better results than 32-bit \float on average of 51\% cases. Including all the case in which 32-bit \float fails, \posit provide similar or better results in 85\% cases.

\section{Hardware Results}
\label{sec:hw-results}

\begin{scriptsize}
\begin{table*}[!t]
    \small
    \begin{center}
    \begin{tabular}{|l|c|c|c|c|c|c|}
    \hline
        \multirow{2}{*}{\textbf{Module}} 
       & \multicolumn{2}{c|}{\textbf{Posit(es=2)}}
       & \multicolumn{2}{c|}{\textbf{Posit(es=3)}}
       & \multicolumn{2}{c|}{\textbf{Posit(es=2,3)}} \\ \cline{2-7}
       & Slice LUTs  & Slice Registers
       & Slice LUTs  & Slice Registers
       & Slice LUTs  & Slice Registers \\  \hline \hline
       
      Fused Multiply-Add & 1128    & 416 &    1154 &    415 & 1176 &    424 \\ \hline
      Division & 281 &    270    & 303 &    237 & 279 &    271  \\ \hline
      Square root & 138    & 145 &    135 &    143 & 138 &    147 \\ \hline
      Integer to Posit & 131 &    37 &    128 &    36 & 131 & 37 \\ \hline
      Posit to Integer  & 163 &    39 &    161 &    39 & 165 &    39\\ \hline
      Sign Injection & 18 &    0 &    18 &    0 & 18 &    0 \\ \hline
      Classify & 2 &    0 &    2 &    0 & 2     & 0 \\ \hline
      Decode Posit & 554 &    0 &    555 &    0 & 684     & 0 \\ \hline
      Encode Posit &  306 &    0 &    307 & 0 & 371 &    0 \\ \hline
      Glue Logic & 307 &    284 &    281 &    311 & 543 &    376 \\ \hline
      \hline
      \textbf{Total} & {\bf 3028}    & {\bf 1191} &    {\bf 3044} &    {\bf 1181} & {\bf 3507} &    {\bf 1294} \\ \hline
    \end{tabular}
    \caption{FPGA Synthesis: Module-wise Slice LUTs and Slice Registers with \posit FPU of {\em es=2}, {\em es=3} and \emph{es=2,3} for 32-bit posit size at 100 MHz}
    \label{tab:fpga_synth}
    \end{center}  
    \end{table*}
\end{scriptsize}

This section presents the synthesis results of parameterized \posit unit for {\em (ps=32,es=2)} and {\em (ps=32,es=3)} instances. Also, we present the overheads for dynamic switching in this section. We have synthesized our designs for an Artix-7 FPGA (xc7a100tcsg324-1) device using Xilinx Vivado 2018.3. We chose to constraint the designs to operate at 100 MHz as most open-source RISC-V cores operate at this frequency for the same FPGA target. It is to be noted that the results presented in this section are independent to the integration choices mentioned in Section-\ref{sec:core-integration}.

Table-\ref{tab:fpga_synth} shows the LUT and slice-register utilisation on a modular basis. For this table, each module has been synthesized separately using the \emph{Default Strategy} of Vivado with re-timing enabled. It should be noted that there is no separate module for dynamic switching as the entire operation is achieved through the decoding and encoding modules itself.  Due to the presence of run-length encoding in \posit, the encoding and decoding modules pose a strong design challenge and thus consume considerable resources.

One can observe from Table-\ref{tab:fpga_synth} that dynamic switching support
requires 15\% more LUTs and 8\% more registers than the base {\emph es=2} implementation.


 Table-\ref{tab:cycle_count} shows the number of cycles taken by different floating-point instructions in the \posit based RISC-V core. We have pipelined the FMA module into 8 stages and to operate at 100 MHz. Since addition and subtraction operation do not need to multiply the fraction, they skip the product stage and thus require 6 cycles to complete. Similarly, multiplication result comes out of the pipeline as soon as the product is calculated, taking a total of 6 cycles. Division and square root operation use iterative non-restoring approaches to calculate quotient and remainder. Hence the number of cycles required is proportional to fraction length. Conversion operations (integer to \posit and \posit to integer) either involves decoding or encoding of the number and they require 3 cycles to complete the operation. Sign injection, move and classify instructions complete in 1 cycle. Switching the {\em es} value involves decoding and encoding logic and hence require 4 cycles.

\begin{scriptsize}
\begin{table}
    \small
    \begin{center}
        \begin{tabular}{|p{6.5cm}|C{1cm}|}
      \hline
      \textbf{Instructions} & \textbf{Cycles} \\ \hline \hline
      FMADD.S, FMSUB.S, FNMSUB.S, FNMADD.S  & 8 \\ \hline
      FADD.S, FSUB.S & 6 \\ \hline
      FMUL.S & 6 \\ \hline
      FDIV.S & 20 \\  \hline
      FSQRT.S & 32  \\  \hline
      FCVT.W.S, FCVT.WU.S & 3 \\  \hline
      FCVT.S.W, FCVT.S.WU & 3 \\  \hline
      FMIN.S, FMAX.S, FEQ.S, FLT.S, FLE.S & 1 \\ \hline
      FSGNJ.S, FSGNJN.S, FSGNJX.S & 1 \\ \hline
      FMV.X.W, FMV.W.X & 1 \\ \hline
      FCLASS.S & 1 \\ \hline
      FCVT.ES & 4 \\ \hline
      
    \end{tabular}
    \caption{No. of cycles required, by each instruction in RV32F standard extension of RISC-V spec in \posit FPU (\emph{es=2,3}) for 32-bit posit size at 100 MHz}
    \label{tab:cycle_count}
    \end{center}
\end{table}
\end{scriptsize}
\section{Related Work}
\label{sec:related_work}

This section provides a comparison of the contributions of this paper with related works in 
literature.


Some of the works in literature have failed in complying their implementations with the
\posit spec. For example, the works in~\cite{date, iscas} present hardware units
for addition, subtraction and multiplication of \posit numbers but do not provide the necessary rounding mode (round-to-nearest with tie-to-even) support. The \posit FPU
presented in this paper supports the rounding mode and complies exactly with the \posit standard. 

There exist a few works, like~\cite{ipdps, iccd}, which try to compare their \posit designs with \float based designs to claim that \posit is a suitable option to replace \float. In contrast to this, our work provides a solution where \posit and \float FPUs can co-exist on the same chip and enable users to choose either for their applications. In addition to this argument, we also believe that comparing hardware overheads of \posit and \float is not fair since \float allows a large amount of flexibility in implementation choices (e.g., supported rounding modes, support for subnormals, nan-propagation, etc.) and choosing a fair design point for comparison with \posit is difficult. The comparison becomes
even more challenging when one accounts for circuit optimizations specific to each
paradigm.

One of the crucial contributions missing in all the above-cited works (including~\cite{sigproc, pacogen}) is
the fact that none of the proposed FPUs are feature-complete to be integrated with a general-purpose processor. In this paper, we not only propose how a \posit based
FPU can be integrated with a RISC-V core but also present a few applications which truly capture the benefits of the \posit over \float.

Works like~\cite{iccd, pacogen}  have proposed a parameterized design of \posit unit which can be used to generate hardware for any {\em ps} and {\em es} values. 
In our work, along with this, we go one step further and provide the capability to support multiple {\em es} modes within the same hardware unit, thereby catering to applications from a wide range of domains. This, to the best of our knowledge, has not been proposed by any previous work in literature.

Our work not only provides implementation details of \posit arithmetic but also captures the difference between \posit and \float arithmetic, which support the claim of \posit being more efficient than \float.

\section{Conclusion}
\label{sec:conclusion}

In this paper, we propose the first parameterized \posit based FPU designed in BSV and integrate with a RISC-V compliant core. The paper provides
a concrete path on how the RISC-V 'F' extension should be extended or modified to support \posit. We also present an alternate methodology of exploiting the \emph{custom} space of RISC-V ISA to integrate the \posit FPU as an accelerator with any RISC-V core supporting a RoCC like interface. The latter strategy thus enables a RISC-V core to support both \float and \posit on the same chip allowing
applications to choose one over the other based on their requirements. The paper
further goes on to highlight the specific differences between \posit and \float in regards to supporting the various 'F' extension instructions of RISC-V. This analysis leads us to conclude that \posit simplifies floating-point arithmetic
significantly as compared to \float. The paper also proposes a novel idea
of supporting multiple \emph{es} values within the same hardware unit with dynamic switching capabilities through minimal overheads. The paper also demonstrates how applications can be ported to a RISC-V core with \posit support in the absence of appropriate compiler support. Scrutiny of these
applications further emphasize the benefits and claims made by \posit over \float. Through FPGA prototyping, we show that an instance of the proposed
\posit FPU consumes around 3.5K LUTs and 1.3K slice registers on a 7-series FPGA.

\bibliographystyle{IEEEtran}
\bibliography{main.bbl}

\begin{thebibliography}{10}
\providecommand{\url}[1]{#1}
\csname url@samestyle\endcsname
\providecommand{\newblock}{\relax}
\providecommand{\bibinfo}[2]{#2}
\providecommand{\BIBentrySTDinterwordspacing}{\spaceskip=0pt\relax}
\providecommand{\BIBentryALTinterwordstretchfactor}{4}
\providecommand{\BIBentryALTinterwordspacing}{\spaceskip=\fontdimen2\font plus
\BIBentryALTinterwordstretchfactor\fontdimen3\font minus
  \fontdimen4\font\relax}
\providecommand{\BIBforeignlanguage}[2]{{%
\expandafter\ifx\csname l@#1\endcsname\relax
\typeout{** WARNING: IEEEtran.bst: No hyphenation pattern has been}%
\typeout{** loaded for the language `#1'. Using the pattern for}%
\typeout{** the default language instead.}%
\else
\language=\csname l@#1\endcsname
\fi
#2}}
\providecommand{\BIBdecl}{\relax}
\BIBdecl

\bibitem{moore}
G.~E.~Moore, ``Cramming more components onto integrated circuits, reprinted
  from electronics, volume 38, number 8, april 19, 1965, pp.114 ff,''
  \emph{Solid-State Circuits Newsletter, IEEE}, vol.~11, pp. 33 -- 35, 10 2006.

\bibitem{dennard}
R.~H. {Dennard}, F.~H. {Gaensslen}, V.~L. {Rideout}, E.~{Bassous}, and A.~R.
  {LeBlanc}, ``Design of ion-implanted mosfet's with very small physical
  dimensions,'' \emph{IEEE Journal of Solid-State Circuits}, vol.~9, no.~5, pp.
  256--268, Oct 1974.

\bibitem{ieee754}
``Ieee standard for floating-point arithmetic,'' \emph{IEEE Std 754-2008}, pp.
  1--70, Aug 2008.

\bibitem{luca}
G.~Tagliavini, S.~Mach, D.~Rossi, A.~Marongiu, and L.~Benini, ``A
  transprecision floating-point platform for ultra-low power computing,''
  \emph{2018 Design, Automation and Test in Europe Conference and Exhibition
  (DATE)}, pp. 1051--1056, 2018.

\bibitem{fang}
\BIBentryALTinterwordspacing
X.~Fang and M.~Leeser, ``Open-source variable-precision floating-point library
  for major commercial fpgas,'' \emph{ACM Trans. Reconfigurable Technol.
  Syst.}, vol.~9, no.~3, pp. 20:1--20:17, Jul. 2016. [Online]. Available:
  \url{http://doi.acm.org/10.1145/2851507}
\BIBentrySTDinterwordspacing

\bibitem{whitehead}
N.~Whitehead and A.~Fit-florea, ``Precision \& performance: Floating point and
  ieee 754 compliance for nvidia gpus.''

\bibitem{software_bugs}
W.~Eric~Wong, X.~Li, P.~A.~Laplante, and M.~Siok, ``Be more familiar with our
  enemies and pave the way forward: A review of the roles bugs played in
  software failures,'' \emph{Journal of Systems and Software}, vol. 133, 06
  2017.

\bibitem{posit}
\BIBentryALTinterwordspacing
Gustafson and Yonemoto, ``Beating floating point at its own game: Posit
  arithmetic,'' \emph{Supercomput. Front. Innov.: Int. J.}, vol.~4, no.~2, pp.
  71--86, Jun. 2017. [Online]. Available:
  \url{https://doi.org/10.14529/jsfi170206}
\BIBentrySTDinterwordspacing

\bibitem{date}
M.~K. {Jaiswal} and H.~K.~. {So}, ``Universal number posit arithmetic generator
  on fpga,'' in \emph{2018 Design, Automation and Test in Europe Conference and
  Exhibition (DATE)}, March 2018, pp. 1159--1162.

\bibitem{iscas}
------, ``Architecture generator for type-3 unum posit adder/subtractor,'' in
  \emph{2018 IEEE International Symposium on Circuits and Systems (ISCAS)}, May
  2018, pp. 1--5.

\bibitem{conga}
\BIBentryALTinterwordspacing
Z.~Leh\'{o}czky, A.~Retzler, R.~T\'{o}th, A.~Szab\'{o}, B.~Farkas, and
  K.~Somogyi, ``High-level .net software implementations of unum type i and
  posit with simultaneous fpga implementation using hastlayer,'' in
  \emph{Proceedings of the Conference for Next Generation Arithmetic}, ser.
  CoNGA '18.\hskip 1em plus 0.5em minus 0.4em\relax New York, NY, USA: ACM,
  2018, pp. 4:1--4:7. [Online]. Available:
  \url{http://doi.acm.org/10.1145/3190339.3190343}
\BIBentrySTDinterwordspacing

\bibitem{ipdps}
A.~{Podobas} and S.~{Matsuoka}, ``Hardware implementation of posits and their
  application in fpgas,'' in \emph{2018 IEEE International Parallel and
  Distributed Processing Symposium Workshops (IPDPSW)}, May 2018, pp. 138--145.

\bibitem{sigproc}
J.~Hou, Y.~Zhu, S.~Du, and S.~Song, ``Enhancing accuracy and dynamic range of
  scientific data analytics by implementing posit arithmetic on fpga,''
  \emph{Journal of Signal Processing Systems}, 11 2018.

\bibitem{iccd}
R.~Chaurasiya, J.~Gustafson, R.~Shrestha, J.~Neudorfer, S.~Nambiar, K.~Niyogi,
  F.~Merchant, and R.~Leupers, ``Parameterized posit arithmetic hardware
  generator,'' 10 2018, pp. 334--341.

\bibitem{pacogen}
M.~K. {Jaiswal} and H.~K.~. {So}, ``Pacogen: A hardware posit arithmetic core
  generator,'' \emph{IEEE Access}, vol.~7, pp. 74\,586--74\,601, 2019.

\bibitem{riscv}
A.~Waterman, Y.~Lee, D.~A. Patterson, and K.~Asanovi, ``The risc-v instruction
  set manual. volume 1: User-level isa, version 2.0,'' 2014.

\bibitem{shakti}
\BIBentryALTinterwordspacing
N.~Gala, A.~Menon, R.~Bodduna, G.~S. Madhusudan, and V.~Kamakoti, ``Shakti
  processors: An open-source hardware initiative,'' in \emph{Proceedings of the
  2016 29th International Conference on VLSI Design and 2016 15th International
  Conference on Embedded Systems (VLSID)}, ser. VLSID '16.\hskip 1em plus 0.5em
  minus 0.4em\relax Washington, DC, USA: IEEE Computer Society, 2016, pp. 7--8.
  [Online]. Available: \url{http://dx.doi.org/10.1109/VLSID.2016.130}
\BIBentrySTDinterwordspacing

\bibitem{rocket}
\BIBentryALTinterwordspacing
K.~Asanović, R.~Avizienis, J.~Bachrach, S.~Beamer, D.~Biancolin, C.~Celio,
  H.~Cook, D.~Dabbelt, J.~Hauser, A.~Izraelevitz, S.~Karandikar, B.~Keller,
  D.~Kim, J.~Koenig, Y.~Lee, E.~Love, M.~Maas, A.~Magyar, H.~Mao, M.~Moreto,
  A.~Ou, D.~A. Patterson, B.~Richards, C.~Schmidt, S.~Twigg, H.~Vo, and
  A.~Waterman, ``The rocket chip generator,'' EECS Department, University of
  California, Berkeley, Tech. Rep. UCB/EECS-2016-17, Apr 2016. [Online].
  Available:
  \url{http://www2.eecs.berkeley.edu/Pubs/TechRpts/2016/EECS-2016-17.html}
\BIBentrySTDinterwordspacing

\bibitem{lowrisc}
``{The lowRISC project},'' \url{https://www.lowrisc.org/}.

\bibitem{ariane}
D.~{Rossi}, F.~{Conti}, A.~{Marongiu}, A.~{Pullini}, I.~{Loi}, M.~{Gautschi},
  G.~{Tagliavini}, A.~{Capotondi}, P.~{Flatresse}, and L.~{Benini}, ``Pulp: A
  parallel ultra low power platform for next generation iot applications,'' in
  \emph{2015 IEEE Hot Chips 27 Symposium (HCS)}, Aug 2015, pp. 1--39.

\bibitem{swerv}
W.~Digital, ``{RISC-V SweRV Core},''
  \url{https://blog.westerndigital.com/risc-v-swerv-core-open-source/}.

\bibitem{piccolo}
Bluespec, ``{Open-source RISC-V CPUs},''
  \url{https://github.com/bluespec/Piccolo}.

\bibitem{rocc}
C.~Yarp, ``{An Introduction to the Rocket Custom Coprocessor Interface},''
  \url{http://www-inst.eecs.berkeley.edu/~cs250/sp16/disc/Disc02.pdf}, accessed
  on 10 November 2017.

\bibitem{bluespec}
``{Bluespec Inc. Bluespec System Verilog},'' \url{https://bluespec.com/}.

\bibitem{softposit}
``Softposit library,'' \url{https://gitlab.com/cerlane/SoftPosit}.

\bibitem{calligo}
``{Posit Numeric Unit (PNU) Implementation by Calligo Technologies},''
  \url{https://posithub.org/conga/2018/docs/9-Calligo-Technologies.pdf}.

\bibitem{vividsparks}
``{Posit Implementation by VividSparks},'' \url{http://vivid-sparks.com/}.

\bibitem{confidence_interval}
``Confidence interval,''
  \url{http://hamelg.blogspot.com/2015/11/python-for-data-analysis-part-23-point.html}.

\bibitem{iris}
R.~A. Fisher, ``The use of multiple measurements in taxonomic problems,''
  \emph{Annals of Eugenics}, vol.~7, no.~7, pp. 179--188, 1936.

\bibitem{wisconsin}
N.~Street, W.~H.~Wolberg, and O.~L~Mangasarian, ``Nuclear feature extraction
  for breast tumor diagnosis,'' \emph{Proc. Soc. Photo-Opt. Inst. Eng.}, vol.
  1993, 01 1999.

\bibitem{ecoli}
P.~Horton and K.~Nakai, ``A probabilistic classification system for predicting
  the cellular localization sites of proteins,'' \emph{Proceedings / ...
  International Conference on Intelligent Systems for Molecular Biology ; ISMB.
  International Conference on Intelligent Systems for Molecular Biology},
  vol.~4, pp. 109--15, 02 1996.

\bibitem{scikit-learn}
F.~Pedregosa, G.~Varoquaux, A.~Gramfort, V.~Michel, B.~Thirion, O.~Grisel,
  M.~Blondel, P.~Prettenhofer, R.~Weiss, V.~Dubourg, J.~Vanderplas, A.~Passos,
  D.~Cournapeau, M.~Brucher, M.~Perrot, and E.~Duchesnay, ``Scikit-learn:
  Machine learning in {P}ython,'' \emph{Journal of Machine Learning Research},
  vol.~12, pp. 2825--2830, 2011.

\end{thebibliography}

\end{document}